\makeatletter \providecommand{\@LN}[2]{} \makeatother
\documentclass[aps,pra,twocolumn,linenumbers,citeautoscript,floatfix]{revtex4-1}
\pdfoutput=1 \usepackage{graphicx} \usepackage{amssymb}
\usepackage{amsmath}
\usepackage[usenames,dvipsnames,svgnames,table]{xcolor}
\usepackage{bm}
\usepackage{mathdots}
\usepackage{float}

\begin{document}

\nolinenumbers
\setlength\linenumbersep{6pt}

\makeatletter \def\imod#1{\allowbreak\mkern10mu({\operator@font
    mod}\,\,#1)} \makeatother

\newcommand{\eqn}[1]{(\ref{eq.#1})}
\newcommand{\dotprod}{{\scriptscriptstyle \stackrel{\bullet}{{}}}}
\newcommand{\bra}[1]{\mbox{$\left\langle #1 \right|$}}
\newcommand{\ket}[1]{\mbox{$\left| #1 \right\rangle$}}
\newcommand{\bl}{{\mbox{\rm \scriptsize B}}}
\newcommand{\blket}[1]{\mbox{$\left| #1 \right\rangle_\bl$}}
\newcommand{\braket}[2]{\mbox{$\left\langle #1 | #2\right\rangle$}}
\newcommand{\av}[1]{\mbox{$\left| #1 \right|$}}
\newcommand{\bk}[1]{\mbox{$\left\langle #1 \right\rangle$}}
\newcommand{\mom}[2]{\bk{#1}_{\scriptscriptstyle #2}}
\newcommand{\osc}{{\mbox{\rm \scriptsize osc}}}
\newcommand{\tot}{{\mbox{\rm \scriptsize tot}}}
\newcommand{\total}{{\mbox{\rm \scriptsize total}}}
\newcommand{\op}[1]{\mathsf{#1}} \newcommand{\tts}{{\hspace{.1em}}}
\newcommand{\lga}{{\mbox{\rm \scriptsize LGA}}}
\newcommand{\swap}{{\mbox{\rm \scriptsize swap}}}
\newcommand{\ground}{{\mbox{\rm \scriptsize ground}}}
\newcommand{\bound}{{\mbox{\rm \scriptsize bound}}}
\newcommand{\cycle}{{\mbox{\rm \scriptsize cycle}}}
\newcommand{\orbit}{{\mbox{\rm \scriptsize orbit}}}
\newcommand{\interval}{{\mbox{\rm \scriptsize interval}}}
\newcommand{\diff}{{\mbox{\rm \scriptsize different}}}
\newcommand{\eff}{{\mbox{\rm \scriptsize eff}}}
\newcommand{\freelymoving}{{\mbox{\rm \scriptsize freely moving}}}
\newcommand{\massless}{{\mbox{\rm \scriptsize massless particles}}}
\newcommand{\light}{{\mbox{\rm \scriptsize light}}}
\newcommand{\particle}{{\mbox{\rm \scriptsize particle}}}
\newcommand{\particles}{{\mbox{\rm \scriptsize particles}}}
\newcommand{\classical}{{\mbox{\rm \scriptsize classical}}}
\newcommand{\internal}{{\mbox{\rm \scriptsize internal}}}
\newcommand{\initial}{{\mbox{\rm \scriptsize initial}}}
\newcommand{\midpoint}{{\mbox{\rm \scriptsize midpoint}}}
\newcommand{\average}{{\mbox{\rm \scriptsize average}}}
\newcommand{\nonrel}{{\mbox{\rm \scriptsize non-rel}}}
\newcommand{\rest}{{\mbox{\rm \scriptsize rest}}}
\newcommand{\twoparticles}{{\mbox{\rm \scriptsize 2p}}}
\newcommand{\ideal}{{\mbox{\rm \scriptsize ideal}}}
\newcommand{\block}{{\mbox{\rm \scriptsize block}}}
\newcommand{\blockchange}{{\mbox{\rm \scriptsize block-change}}}
\newcommand{\statechange}{{\mbox{\rm \scriptsize state-change}}}
\newcommand{\changed}{{\mbox{\rm \scriptsize changed}}}
\newcommand{\unchanged}{{\mbox{\rm \scriptsize unchanged}}}
\newcommand{\change}{{\mbox{\rm \scriptsize change}}}
\newcommand{\nochange}{{\mbox{\rm \scriptsize same}}}
\newcommand{\kinetic}{{\mbox{\rm \scriptsize kinetic}}}
\newcommand{\potential}{{\mbox{\rm \scriptsize potential}}}
\newcommand{\motion}{{\mbox{\rm \scriptsize motion}}}
\newcommand{\local}{{\mbox{\rm \scriptsize local}}}
\newcommand{\hop}{{\mbox{\rm \scriptsize hop}}}
\newcommand{\locations}{{\mbox{\rm \scriptsize locations}}}
\newcommand{\even}{{\mbox{\rm \scriptsize even}}}
\newcommand{\odd}{{\mbox{\rm \scriptsize odd}}}
\newcommand{\rt}{{\mbox{\rm \scriptsize R}}}
\newcommand{\lt}{{\mbox{\rm \scriptsize L}}}
\newcommand{\shift}{{\mbox{\rm \scriptsize shift}}}
\newcommand{\ex}{{\mbox{\rm \scriptsize ex}}}
\newcommand{\D}[2]{\frac{\partial #2}{\partial #1}}
\newcommand{\pp}{{\mbox{\tt \scriptsize /}}}
\newcommand{\mm}{{\mbox{\tt \scriptsize \backslash}}}
\newcommand{\x}{{\tilde{x}}}
\newcommand{\y}{{\tilde{y}}}
\newcommand{\z}{{\tilde{z}}}
\newcommand{\smax}{{\mbox{\rm \scriptsize max}}}
\newcommand{\smin}{{\mbox{\rm \scriptsize min}}}
\newcommand{\spart}{{\mbox{\rm \scriptsize part}}}
\newcommand{\savg}{{\mbox{\rm \scriptsize avg}}}
\newcommand{\sperp}{{\scriptscriptstyle \perp}}
\newcommand{\taud}{{\tau_{\scriptscriptstyle \Delta}}}
\newcommand{\Nd}{{N_{\scriptscriptstyle \Delta}}}
\newcommand{\minbw}{{\mbox{\rm \scriptsize MB}}}
\newcommand{\sub}{{\mbox{\em \scriptsize subsystem}}}

\newcommand{\taumin}{\tau_{\smin}}
\newcommand{\taumax}{\tau_{\smax}}
\newcommand{\taustarmin}{\tau_{\smin\,\star}}
\newcommand{\tauavg}{\tau} \newcommand{\lambdamin}{\lambda_{\smin}}
\newcommand{\lambdaavg}{\lambda}
\newcommand{\cc}{c \hskip .05em}

\newcommand{\n}{{\boldsymbol{n}}} \newcommand{\m}{{\boldsymbol{m}}}
\renewcommand{\i}{{\boldsymbol{i}}} 
\renewcommand{\j}{{\boldsymbol{j}}} 
\renewcommand{\k}{{\boldsymbol{k}}} 
\renewcommand\Re{\operatorname{\mathfrak{Re}}}

\newcommand{\sinc}{{\mbox{\rm sinc}}\:}
\newcommand{\sincs}{{\mbox{\rm sinc$^2$}}}

\setlength{\fboxsep}{.1pt} \setlength{\fboxrule}{.1pt}

\title{The finite-state character of physical dynamics}

\begin{abstract}
Finite physical systems have only a finite amount of distinct state.
This finiteness is fundamental in statistical mechanics, where the
maximum number of distinct states compatible with macroscopic
constraints defines {\em entropy}.  Here we show that finiteness of
distinct state is similarly fundamental in ordinary mechanics: {\em
energy} and {\em momentum} are defined by the maximum number of
distinct states possible in a given time or distance.  More generally,
{\em any moment of energy or momentum} bounds distinct states in time
or space. These results generalize both the Nyquist bandwidth-bound on
distinct values in classical signals, and quantum uncertainty bounds.
The new {\em certainty bounds} are achieved by finite-bandwidth
evolutions in which time and space are effectively discrete, including
quantum evolutions that are effectively classical.  Since energy and
momentum count distinct states, they are defined in finite-state
dynamics, and they relate classical mechanics to finite-state
evolution.
\end{abstract}

\author{Norman Margolus}
\affiliation{Massachusetts Institute of Technology, Cambridge MA
  02139. {\tt nhm@mit.edu}}

\maketitle

\definecolor{light-gray}{gray}{0.95}

\section{Introduction}

We live in a world that, like a digital photograph, has only finite
resolution.  This was first recognized in statistical mechanics, when
Planck introduced a finite grain-size $h$ to get a realistic counting
of distinct states \cite{planck,planck2}.  Once it was understood that
$h$ relates all energy and momentum to
waves \cite{deBroglie,schrodinger}, finite resolution was explained as
a property of waves: a tradeoff between range of frequencies superposed
and maximum
localization \cite{bohr,heisenberg,uffink,donoho,max-speed,braunstein,yu,mandelstam,bhatt,luo1,zych,zozor,angulo,g0,chau}.


There is also a tradeoff, in superpositions of waves, between
frequency range and \mbox{\em average} localization.  This is known
in communications theory as the Nyquist bound: a finite bandwidth
signal can carry only a finite number of distinct values per unit
length.  This holds because a finite number of Fourier components
can add up to chosen values at only a finite number of
places \cite{nyquist}.

In this paper, we combine and generalize these tradeoffs.  We count
{\em how many} quantum states can be distinguished from each other
{\em with certainty}, in a finite time or distance, given average
constraints on wavefunction bandwidth. These {\em certainty bounds}
redefine energy and momentum as maximum counts, and challenge the
distinction between continuous and discrete in physics.

To illustrate the connection between bandwidth and distinct quantum
states, consider a free particle moving in one dimension, in a periodic
space of length $L$.  Momentum eigenstates must have a whole number of
oscillations in period $L$, so allowed spatial frequencies $p_n/h$ are
$1/L$ apart.  A wavefunction using $N$ different spatial frequencies
must have at least $N-1$ times this minimal separation, between minimum
and maximum frequencies:
\begin{equation}\label{eq.wl}
      \frac{p_\smax - p_\smin}{h} \ge \frac{N-1}{L}\;.
\end{equation}
This is a bandwidth bound for a superposition of $N$ distinct
energy-momentum eigenstates. It also bounds the total number of
distinct states that can occur as the particle moves a distance $L$:
$N$ distinct eigenstates can add up to at most $N$ distinct
sums. Similar arguments apply to energy and time, for an evolution
periodic in time \cite{emulation}.

More generally, {\em any absolute moment} of energy or momentum is an
average measure of the frequency-width of the wavefunction, and can
play the role that momentum-bandwidth does in \eqn{wl}, determining a
maximum count of distinct states for {\em any portion of any evolution}
with that moment.  For $N=2$ these tradeoffs become minimum uncertainty
relations.  To achieve the maximum count, the wavefunction must use a
finite range of frequencies.  Then the exact evolution can be
interpolated from the state on a discrete set of points in space and
time \cite{emulation,interpolation,kempf1,kempf2,succi,birula,meyer,yepez-path,yepez-gr,yepez,qsim,min-len,mauro,tsang,ideal-energy}.


Perhaps the most interesting moment is average energy above the minimum
possible \cite{max-speed}.  What we call {\em energy} classically,
counts how many distinct states can occur in a unit of time.  How much
change.  How many distinct computational steps.  We can also count just
the distinct states due to overall motion, by comparing energy counts
in rest and non-rest frames.  Surprisingly, motional change is bounded
not by the kinetic energy $E-m c^2$, but by $p v$ instead.  This
difference makes the classical action a count of possible distinct
states.  It also defines an ideal momentum in finite-state
dynamics \cite{ideal-energy}.  Of course, energy also bounds what can
be distinguished experimentally.  For example, using
optics \cite{kok,g1,g2,hall,zwierz}, with $n$ photons of the same
frequency there are at most $2n+1$ distinct phases within one cycle of
oscillation, according to \eqn{eband}.



Below, we first establish energy bounds on the maximum number of states
distinguishable-with-certainty that can occur in a given time.  We then
establish related certainty bounds on overall motion, and discuss
finite distinctness in classical dynamics.  The arguments used are
elementary, and the results are verified numerically.

\section{Distinguishability in time}

For an evolution with period $T$, passing through $N$ distinct
(mutually orthogonal) states at a {\em constant rate}
\cite{max-speed},
\begin{equation}\label{eq.eband}
\frac{2(E-E_0)}{h} \ge \frac{N-1}{T}\;,
\end{equation}
where $E_0$ is the lowest energy eigenvalue used in constructing the
system's state.  The left side is, as in \eqn{wl}, a measure of the
width of an eigenfrequency distribution: twice the average half-width.
The right side is, again, the minimum frequency width for $N$ distinct
states.  We show that \eqn{eband} holds even if the time intervals
between distinct states are unconstrained.  Then, letting $\tauavg$ be
the average time separating consecutive distinct states, \eqn{eband}
becomes
\begin{equation}\label{eq.et}
(E-E_0) \,\tauavg \ge \frac{N-1}{N}\frac{h}{2}\;.
\end{equation}
We show \eqn{et} also holds for a {\em portion of an evolution},
comprising $N$ distinct states with average separation $\tauavg$. For
$N=2$ this becomes the minimum separation bound \cite{max-speed}.  We
provide similar bounds for other moments of energy.

{\bf We formalize our problem as a minimization.}  Consider a
finite-sized isolated system with a time evolution expressed as a
superposition of energy eigenstates:
\begin{equation}\label{eq.kett}
\ket{\psi(t)} = \sum_n a_n e^{- 2\pi i \nu_n t}\ket{E_n}\;,
\end{equation}
with $\nu_n=E_n/h$.  We define a set of average frequency widths
(moments) about a frequency $\alpha$:
\begin{equation}\label{eq.dnu}
\mom{\nu-\alpha}{M} \equiv \Bigl(\sum_n \av{a_n}^2
\,\av{\nu_n-\alpha}^M\Bigr)^{\frac{1}{M}}\;,
\end{equation}
with $M>0$ (other measures of overall width could also be used, {\em
e.g.} \cite{uffink}). If evolution \eqn{kett} passes through a series
of mutually orthogonal states $\ket{\psi(t_k)}$ at times $t_k$, then
\begin{equation}\label{eq.orthogt}
\braket{\psi(t_m)}{\psi(t_k)}=\sum_n \av{a_n}^2 e^{2\pi i
  \nu_n(t_m-t_k)} = \delta_{mk}\;.  
\end{equation}
We seek the minimum frequency widths \eqn{dnu} of states satisfying
the constraints \eqn{orthogt} for any sequence of $N$ distinct
states within a time interval of length $T_N$.  

We assume, {\em without loss of generality}, that all $\nu_n$ are
distinct (in both \eqn{dnu} and
\eqn{orthogt}, coefficients for a repeated $\nu_n$  can be consolidated), and that overall evolution is
periodic with some recurrence-time \cite{qm-recurrence} $T$ that may be
much longer than $T_N$.  Then the discrete spectrum, bound\-ed from
below \cite{verch}, includes {\em at most} all of the frequencies
\begin{equation}\label{eq.nun}
\nu_n=\nu_0 + n/T\;,
\end{equation}
with $n$ a non-negative integer.  These are all of the possible
eigenfrequencies of energy eigenstates that cycle with period $T$,
up to an overall phase.  This spectrum restricts the {\em maximum
period} to be $T$, but evolution can repeat more than once in this
time.  For $T$ sufficiently large \eqn{nun} approaches a continuous
spectrum, allowing us to minimize over the union of all possible
discrete spectra.

\begin{figure}[t]{ \hfill \mbox{
  \includegraphics[height=1.79in]{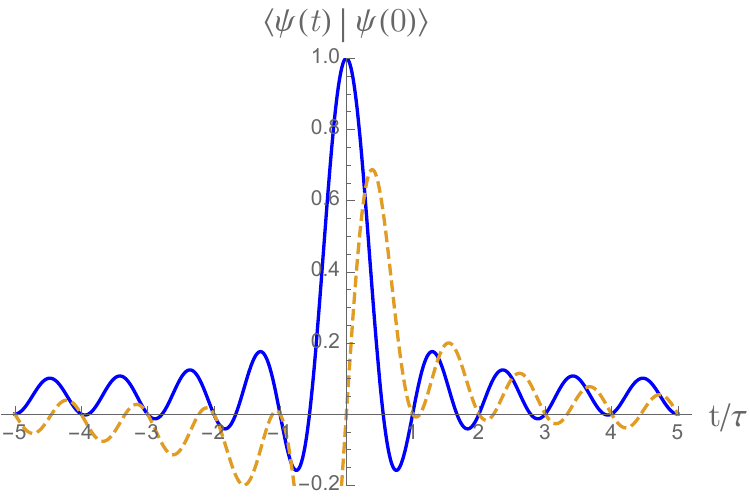}}\hfill} 
  \caption{{\it A
  periodic evolution with $N$ distinct states $\tauavg$ apart}
  (solid real, dashed imaginary, depicted for $N=10$).  Only a
  discrete set of frequencies fit the period: all are allowed in
  the minimization.  An equally weighted superposition
  $\ket{\psi(t)}$ of $N$ consecutive frequencies is the narrowest
  that gives $N$ distinct states in time. Centered on $\alpha$, it
  minimizes all $\tauavg\, \mom{\nu-\alpha}{M}$.}
\label{fig.overlap}
\end{figure}

{\bf We first consider an evolution with a constant rate of
distinct change.} If $N>1$ distinct states have equal separations
$\tauavg$ within period $T=N\tauavg$, then $t_m = m\tauavg$ and
from \eqn{orthogt} and \eqn{nun},
\begin{equation}\label{eq.nsum}
\braket{\psi(t_{k+m})}{\psi(t_k)}= e^{2\pi i \nu_0 t_m}
\sum_{n=0}^\infty \av{a_n}^2 e^{2\pi i n m/N}\;.
\end{equation}
There are only $N$ distinct phases in the sum \eqn{nsum}, so we can
minimize all $\mom{\nu-\alpha}{M}$ for a given $\alpha$ by using a
set of $N$ consecutive $\nu_n$'s, centered as closely as possible on
$\alpha$: we get the same orthogonality times in \eqn{nsum} with
smaller width \eqn{dnu} by setting each $\av{a_n}^2$ outside the
set to 0, and transferring its weight to the equivalent phase
within the set. Then, since
$\braket{\psi(t_{k+m})}{\psi(t_k)}=\delta_{m0}$, the $N$
consecutive non-zero $\av{a_n}^2$ are just the discrete Fourier
transform of a Kronecker delta impulse, and so they all equal
$1/N$.  Thus all $\mom{\nu-\alpha}{M}$ are minimized by an equal
superposition with minimum band\-width for $N$ distinct states
(illustrated in Figure~\ref{fig.overlap}), so the dimensionless
product
\begin{equation}\label{eq.fn}
\tauavg\,\mom{\nu-\alpha}{M} \;\ge\; f_\alpha(M,N)\;
\end{equation}
for some $f_\alpha(M,N)$ defined by the minimizing state.  For example,
if $\alpha=\nu_0$, the closest to centering a minimum bandwidth state
on $\alpha$ is for $\nu_0$ to be the lowest frequency.  Then equality
in \eqn{fn} requires $N$ equal $\av{a_n}^2$ in \eqn{dnu}, and
\begin{equation}\label{eq.fnu0}
f_{\nu_0}(M,N) = N^{-(1+\frac{1}{M})}
\,(\textstyle{{\sum_{n=0}^{N-1} n^M}})^{\frac{1}{M}}\;.
\end{equation}
For $M\ge 1$ this ranges from $1/4$ to $1$.  $f_{\nu_0}(1,N)$
gives \eqn{et}.

Similarly, if $\alpha$ is the midpoint of $N$ consecutive frequencies
$\nu_n$, a minimum bandwidth state can be exactly centered on the mean
frequency $\bar\nu=\alpha$, and
\begin{equation}\label{eq.fnubar}
f_{\bar\nu}(M,N) = N^{-(1+\frac{1}{M})}
\,(\textstyle{{\sum_{n=0}^{N-1}
    \av{n-\frac{N-1}{2}}^M}})^{\frac{1}{M}}\;.
\end{equation}
For $M\ge1$ this ranges from $2/9$ to $1/2$, and is the smallest
achievable bound: no other $\alpha$ gives a smaller bound.  For
$0<M<1$, however, $\alpha=\nu_n$ (an eigenfrequency) is better for even
$N$ (even though a minimum bandwidth state can't be exactly centered on
$\alpha$ in this case).  Excluding $f_{\bar\nu}$ for $M<2$, both
$f_{\nu_0}$ and $f_{\bar\nu}$ strictly increase with $N$.



{\bf Now consider an evolution with a constant rate portion}.
Suppose there are $N$ distinct states, spaced $\tauavg$ apart,
within an interval $T_N$.  To find the minimum of
$\tauavg\, \mom{\nu-\alpha}{M}$ we assume evolution outside of
$T_N$ puts no constraints on the minimization problem: it adds no
orthogonality constraints, and the maximum period $T$ is unbounded
so all frequencies are allowed in \eqn{nun}.

\begin{figure}[t]{ \hfill \mbox{
    \includegraphics[height=1.79in]{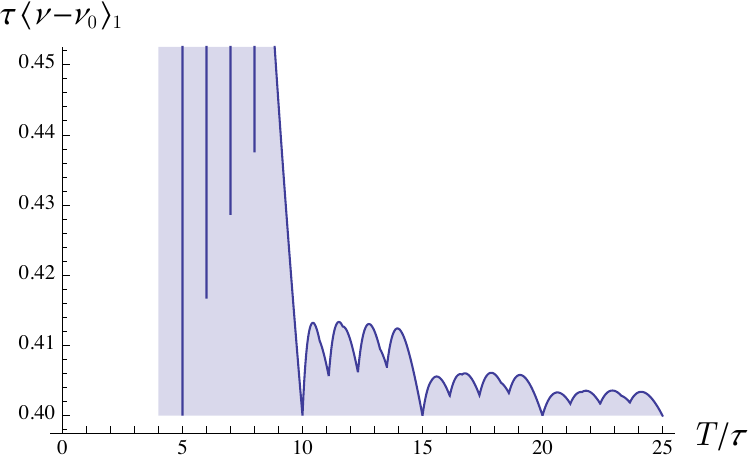}}\hfill} 
    \caption{{\it
     Minimum of $\tauavg\mom{\nu-\nu_0}{1}$ for an evolution with
     maximum period $T$ that includes $N=5$ distinct states,
     $\tauavg$ apart.}  Each choice of $T$ constrains the frequency
     spectrum, and the corresponding minimum is determined
     numerically.  The minimum for $T=N\tauavg$ (bottom of shaded
     area) recurs, and is the minimum for $T\to\infty$, the case of
     an unconstrained spectrum.}
\label{fig.epart}
\end{figure}

We find, in general, that the optimal evolution containing $T_N$
repeats with period $N\tauavg$, and so the bounds are again
$f_\alpha(M,N)$.  We see this in the example of Figure~\ref{fig.epart},
plotting the minimum of $\tauavg \mom{\nu-\nu_0}{1}$ with $N=5$ for
different $T$ (for numerical methods see Appendix~\ref{sec.tests}).
The global minimum recurs whenever $T$ is an integer multiple of
$N\tauavg$, so the bound for $T=N\tauavg$ holds for $T\to\infty$.

The general behavior is clear for large $M$: the minimum bandwidth
state is the minimizing state, since $\tauavg \mom{\nu-\alpha}{\infty}$
{\em is} the (dimensionless) bandwidth.  Minimum bandwidth requires
repetition with period $N\tauavg$, since otherwise there are too many
constraints \eqn{orthogt} to satisfy.  Similarly for large $N$, the
constant-rate bounds $f_\alpha(M,N)$ apply as long as $f_\alpha(M,N)$
increases monotonically with $N$, since the limit $N\to\infty$ is also
the limit $T=N\tauavg\to\infty$.

The situation is much the same for small $M$ and $N$.  If we take the
limit $T\to\infty$ with $T$ an integer multiple of $\tauavg$ then,
since $t_m=m\tauavg$, for each $T$ there are only $T/\tauavg$ distinct
phases in \eqn{orthogt}, and so only a finite bandwidth $1/\tauavg$ is
relevant to the minimization in the limit.  We surveyed ten thousand
cases numerically (some illustrated in
Figures~\ref{fig.epart-min-survey}--\ref{fig.epart-asymp} of
Appendix~\ref{sec.tests}) and found that the minimizing bandwidth is
slightly smaller: $(N-1)/N\tauavg$, the minimum possible (which
requires repetition with period $N\tauavg$).  The only exceptions were
some moments about $\bar\nu$ with $M<2$ and $N$ even, where the
minimizing bandwidth was $1/\tauavg$.  For these moments, the increase
of $f_{\bar\nu}(M,N)$ with $N$ is non-monatonic: it decreases going
from even to odd.  Thus allowing one more distinct state (or one more
frequency) when $N$ is even decreases the bound.

Equal separation is not an assumption for $N=2$, since there is only
one separation.  Known bounds on orthogonality time agree with with
$f_\alpha(M,2)$ \cite{max-speed,luo1,zych,mandelstam,bhatt}.



{\bf Finally, if equal separation is optimal, \mbox{constant} rate
bounds hold with $\bm\tauavg$ the {\em average separation.}}
Intuitively, inequality requires some separations to be smaller than
average, and it is the smallest separations in an evolution that
require the largest frequency widths, hence equality is optimal.  More
formally, consider an evolution with $N$ distinct states in period $T$.
There are $N$ different intervals $T_N <T$ that encompass all $N$ of
the distinct states, each omitting one separation from $T$.  Since
there is a positive minimum value for the dimensionless product
$T_N\,\mom{\nu-\alpha}{M}$, the smallest $T_N$ requires the largest
$\mom{\nu-\alpha}{M}$. Thus the best we can do is make all $T_N$ equal,
and hence all separations equal.


\begin{figure}[t]{ \hfill \mbox{
    \includegraphics[height=1.75in]{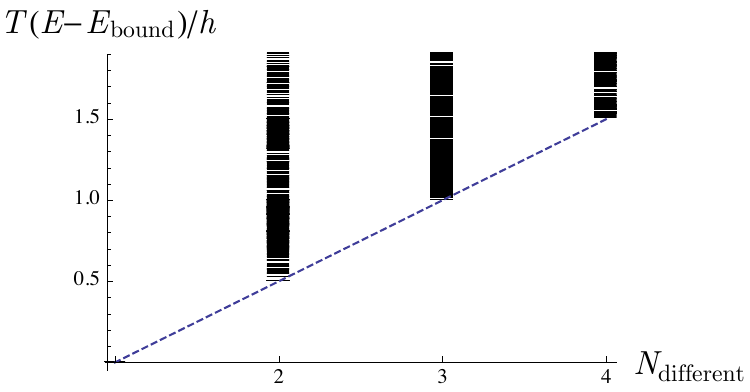}}\hfill} 
    \caption{{\it
     For each of 12,000 sets of separations between distinct states, we
     compare the minimum of $E$ with the minimum $E_\bound$ possible if
     all separations were equal.}  Each evolution is periodic with
     period $T$, and we group them based on the number $N_\diff$ of
     different separation lengths between consecutive distinct states.
     $E>E_\bound$ unless $N_\diff =1$.}
\label{fig.unequal-times}
\end{figure}

Alternatively, we can observe that the number of
constraints \eqn{orthogt} on the minimization problem grows rapidly as
the number $N_\diff$ of different-length separations required in the
evolution increases, and these additional constraints increase the
minimum even when the lengths are almost exactly equal. This is
illustrated in Figure~\ref{fig.unequal-times}, which compares the
minimum average energy $E$, determined numerically for 12,000 random
sets of separations in periodic evolutions, to the equal-separations
minimum $E_\bound = h(N-1)/2T + E_0$ given by \eqn{eband}.  $E_\bound$
is only achievable if $N_\diff=1$.  The dashed line is approached by
almost-equal separations.  As long as the separations aren't exactly
equal the minimum is altered by a discrete jump for each additional
length: requiring $\braket{\psi(t)}{\psi(0)}=0$ for times arbitrarily
close to equal separation essentially adds a slope$\;=0$ constraint at
equal separations (which results in the dashed line), so average energy
is greater.  Other cases are similar (see
Figures~\ref{fig.almost-equal-scaling}--\ref{fig.almost-equal-overlap}
in Appendix~\ref{sec.tests}).

{\bf The moment bounds are achievable:} exactly for spectra that
include $N$ evenly spaced energy {\mbox eigenvalues}, and approximately
for almost even spacing.  In the macroscopic limit, they are achieved
by states that have a uniform probability density for a range of
energies \cite{max-speed}, which is nearly the case for very
complicated evolutions (see Appendix~\ref{sec.rand-h}).  Moreover,
energy can always be moved to a \mbox{system} where \eqn{et} is
achievable, so {\em average energy is equivalent to a count of possible
distinct states per unit time}.

States that achieve a moment bound have essentially minimal bandwidth,
making continuous evolution an interpolation of a discrete one with the
widest possible spacing between discrete states (see
Appendices~\ref{sec.interpol} and \ref{sec.sampled}).


\section{Distinguishability in space}
For an isolated system in motion, some distinct states can be
attributed to the motion.  We can determine how many by comparing with
the same evolution seen in its rest frame: any extra distinct states
when moving must be due to the motion.  Energy bounds the number of
distinct states in each frame, yielding a bound on motion ({\em
cf.} \cite{deBroglie}): 
\begin{equation}\label{eq.px}
p \,\lambdaavg \ge \frac{N-1}{N}\frac{h}{2}\;.
\end{equation}
Here $p$ is the magnitude of the system's average momentum, and
$\lambdaavg$ is the average separation in space within a sequence of
$N$ states that are distinct due to the motion.  Similar bounds hold
for other moments of momentum.

{\bf We first count macroscopically, in two frames.} Assuming $N\gg 1$,
set $h=2$ and take the energy of flat empty space to be $E_0=0$ in both
frames \cite{verch,volovik}.  Then \eqn{et} becomes $1/\tauavg\le E$,
and energy is the maximum average rate of distinct state change
physically possible.

\begin{figure}[t]{%
$\begin{array}{c}
\underbrace{\includegraphics[width=2.5in]{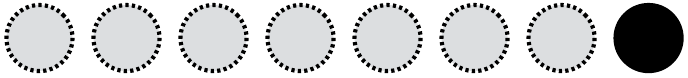}}
\\[.1in] \scalebox{1.4}{$N\le p\Delta x$} \\
\end{array}$}
  \caption{{\it We see extra distinct states of a particle when there
  is relative motion, and we see it as having more than its rest
  energy.}  We can count the extra states based on the extra energy.
  For $N\gg 1$ and using units with $h=2$, maximum distinct states in
  the lab frame is $E\Delta t$, in the rest frame $E_r\Delta t_r$, and
  so the difference $p\Delta x$ is due to overall
  motion.}  \label{fig.particle}
\end{figure}

In the laboratory frame, in a time interval $\Delta t$, an isolated
system evolves through at most $E\Delta t$ distinct states.
Meanwhile, moving at speed $v$, it travels a distance $\Delta
x=v\Delta t$. In the corresponding rest frame evolution, at most
$E_r\Delta t_r$ states are distinct.  The difference, which is a
familiar relativistic quantity
\begin{equation}\label{eq.edt}
E\Delta t - E_r\Delta t_r = p \Delta x\;,
\end{equation}
counts the extra distinct states possible in the frame where there is
overall motion (Figure~\ref{fig.particle}).  Thus $p$ is the extra per
unit distance, agreeing with \eqn{px} for $N\gg 1$.

Dividing \eqn{edt} by $\Delta t$, we see that $E-E_r/\gamma=vp$ bounds
the average rate of motional state change, even at low velocities. This
is slightly surprising, since conventionally the smaller quantity
$E-E_r$ is taken as the energy of motion. Indeed, if we model the
motion of a free particle by treating its rest energy $E_r$ as its
minimum possible energy $E_0$, then \eqn{et} gives $E-E_r$ as the
maximum average rate of motional state change, for $N\gg 1$.  In
general, though, $E_r$ is the average energy of a rest frame dynamics,
so $E-E_r$ is the difference of maximum rates in two different
frames---which is not a rate in either.

{\bf To find precise momentum bounds, consider a constant-speed shift
dynamics.}  With the hamiltonian $\op{H} = v \op{p}_x$, the
wavefunction shifts in the $+x$ direction at speed $v$.  If $\tauavg$
is the average time between distinct states, $\lambdaavg= v\tauavg$ is
the average shift between them.  Since $E_n=v p_n$, we can let
$E_n \tauavg \to p_n \lambdaavg$ in \eqn{fn}, with $\mu_n=p_n/h$ the
spatial frequencies along the direction of motion, giving
\begin{equation}\label{eq.fnp}
\lambdaavg\,\mom{\mu-\alpha}{M} \;\ge\; f_\alpha(M,N)\;.
\end{equation}
This is the general shift-in-space counterpart of the shift-in-time
bound \eqn{fn}. If there is no constraint on the lowest frequency in
the spatial superposition, then the $\alpha=\bar\mu$ bounds apply---the
minimizing state is centered on $\alpha$.


What we seek, however, is more specific: the average separation between
states {\em distinct due to overall motion}.  To find this we need a
wavefunction that represents only overall motion, and nothing of
internal (rest-frame) dynamics.  Thus for a massless particle, we must
require that all momenta along the direction of motion be positive.
Otherwise there will be cancellation of momenta in the superposition,
and part of the energy of the wavepacket will actually be rest energy,
rather than energy of overall motion.  With this restriction the
dynamics is a pure shift so the bounds \eqn{fnp} apply, including the
$\alpha=\mu_\smin$ bounds.

The same conclusions hold for a massive particle.  The moving
wavepacket now has a rest frame, so the only way to avoid seeing {\em
rest-frame changes} in the overall motion is for there to {\em be
none:} a pure shift dynamics.  Thus in all cases, overall motion is
represented by $\op{H} = v \op{p}_x$ with a positive momentum spectrum,
and so $\bk{\op{H}}=v p$.


The bounds \eqn{fnp} are consistent with \eqn{wl}, \eqn{px}, Luo's
bound \cite{luo2} on $\bk{\av{{\op{p}}}}$, and Yu's bound \cite{yu}
$\Delta p \,\lambdamin \ge {h/4}$.  In applying bounds \eqn{fn}
and \eqn{fnp} to computation, note that intermediate results may not be
distinct \cite{jordan}.

\section{Finite classical distinctness}
Although bounds on certainty seem quintessentially quantum mechanical,
finite distinctness of finite-energy physical dynamics is evident even
in the classical realm.


{\bf Macroscopic distinctness is governed by macroscopic energy and
momentum.}  Unlike typical small systems \cite{kok2}, macroscopic
systems traverse a succession of almost perfectly distinct states
as they explore their enormous state spaces: two randomly-chosen
$d$-dimensional normalized states have expected overlap of
$1/\sqrt{d}$, so a sequence of states far enough apart in time to
each be distinct from the next, should all be nearly distinct.

We can investigate how quickly a complicated dynamics reaches distinct
states by studying random dynamics.  For random hamiltonian matrices
(Appendix~\ref{sec.rand-h}), in the limit where the dimension goes to
infinity, with a generic $\psi(0)$ and taking $h=2$ and $E_0=0$, the
overlap $\braket{\psi(t)}{\psi(0)}$ is $2 J_1(\pi Et)/\pi Et$. The
first zero occurs at $t\approx 1.22/E$, close to the bound $\tau \ge
1/E$.  Since the exact dynamics of all the energy in even a tiny
portion of a macroscopic system is so complicated, this may provide at
least a rough idea of the local rate of distinct change: nearly
maximal.


The discretely-distinct character of macroscopic evolution suggests
that finite-state systems should be of fundamental interest in modeling
the classical realm.  Historically, this has been true for modeling
finite entropy in statistical mechanics \cite{ruelle}, but not for
modeling finite energy and momentum in dynamics, where classical
finite-state models have generally been regarded as mere computational
treatments of the ``real'' continuum dynamics
\cite{toffoli-diff,fredkin-dm}.  An exception has been finite-state
lattice models isomorphic to continuum models sampled at integer times
(see Appendix~\ref{sec.cm}). These are closely related to quantum
models with a bounded spectrum (see Appendix~\ref{sec.sampled}).


Macroscopically, if total relativistic energy counts total rate of
distinct change, we can divide this count up into different forms
of energy, and into hierarchies of almost-isolated sets of degrees
of freedom---described by hamiltonians or lagrangians.  Just as the
hamiltonian counts distinct states, so does the lagrangian ({\em
cf.} \cite{brown}).  For example, in a system of particles moving
freely between collisions, $p_i v_i$ counts distinct changes per
unit time due to motion of particle $i$, so the lagrangian $-L =
H-\sum p_i v_i$ counts the changes {\em not} due to particle
motion.



{\bf Classical finite-state models have an ideal energy and momentum.}
From the viewpoint of quantum computation, classical reversible
computation is a special case of what a quantum evolution can do
\cite{bennett-review}.  Classical mechanics doesn't have this
status, because it has an {\em infinite rate} of distinct state
change. Only classical {\em finite-state} dynamics can be recast as
{\em finite-energy} quantum dynamics, with distinct classical
configurations identified with distinct quantum states (see
Appendix~\ref{sec.interpol} and \ref{sec.sampled}).  If we find the
least-energetic realization mathematically possible, no physical
implementation can do better.

A realistic quantum realization is constrained both by certainty bounds
and by relativity.  For example, if a particle travels at speed $v$
through a long sequence of distinct position states $\lambdaavg$~apart,
its minimum possible momentum is $p= h/2\lambdaavg\,$, and the energy
required by distinct motion is $p v$.  If $v<c$ though, total energy
must be larger, since relativistically $E=p v/(v/c)^2>p v$.  We can use
this observation to assign a realistic {\em extensive} ideal-energy to
momentum-conserving lattice models \cite{ideal-energy}.

It might seem surprising that it is, in fact, possible to recast a
classical finite-state dynamics with perfect locality and
determinism, as a quantum hamiltonian dynamics with {\em
continuous} space and time \cite{emulation}.  In this case,
finite-distinctness is encoded in the finiteness of the energy and
momentum of the initial state. The desired finite-state evolution
constitutes a finite set of distinct sample values, which are
continuously interpolated in space and time.  Quantum bounds on
certainty simply reflect finite distinctness in a continuous
description.

{\bf Classical signals obey a version of the bounds.} A classical
signal is like the wavefunction of a scalar particle evolving under
a one-dimensional shift dynamics, $\op{H} = v \op{p}_x$. Any finite
frequency-moment bounds the number $N$ of distinct states in an
interval of the quantum evolution, hence at most $N$ points in the
interval can have values specified independently, by superposing
the distinct states. This generalizes the Nyquist rate
\cite{nyquist} from a bandwidth bound to an any-frequency-moment
bound.







\vspace*{-.8em}
\section{Conclusions}
In the standard quantum description of nature, distinctness is finite
even though time and space are continuous.  There is no contradiction
here, though, because energy and momentum are wave phenomena, and
constraints on their bandwidth impose finite distinctness---just as
they do for continuous classical signals.  In fact, {\em any absolute
moment of a system's energy or momentum bounds the number of distinct
states possible in a given time or distance.}  Since first-moments are
always finite (because average energy and momentum are finite),
distinctness in time and space is always finite.



The moment bounds trade less distinctness in energy or momentum for
more distinctness in time or space.  For a finite system that achieves
a bound, the wavefunction uses only a finite number of energy or
momentum eigenstates, maximally indistinct, and the evolution traverses
an equal number of states that are perfectly distinct in time or space.
Thus the distinct states form a basis for the evolution.  They are
distinct samples of a bandlimited wave, and the states between them are
just interpolation.  This is discrete dynamics in continuous clothing,
and even classical finite-state dynamics can be fit into these clothes.


The fact that {\em every moment} bounds the rate of distinct change in
time or space seems not to be well know; most discussions of the
minimum time for a distinct change single-out just two moments,
discussions of the minimum shift just one.  The fundamental role of
average energy as a conserved distinctness-resource also seems not to
be appreciated: different forms of energy in mechanics identify
different forms of distinctness.  For example, in a
special-relativistic context, $p v$ counts just the distinct states
that are allowed by overall motion.  This motional energy lets us infer
a minimum total relativistic energy by counting distinct changes in
space, enabling the construction of finite-state models of relativistic
systems \cite{ideal-energy}.



As long as energy is finite, the rate of distinct change remains finite
no matter how large a system is or how classical the large-scale
evolution becomes. At a mesoscopic scale we might generally
expect---considering the full dynamics of all energy---that
distinctness in time and space is nearly maximal and the quantumness of
the distinct states is irrelevant.  This suggests that we may be able
to interpret classical mechanics {\em as if the underlying
finitely-distinct evolution were classical}, with classical energy and
momentum governing discreteness in time and space ({\em
cf.} \cite{marsden,bahr}).  This might, for example, allow
informational questions about
gravity \cite{hawking,braun-bh,unruh-info}, modeled as an entropic
force \cite{bek,jacobson1,padman,verlinde,lloyd-geom}, to be studied
classically ({\em cf.} \cite{unruh-sonic,thooft-ca,mottola}).

Finally, if we can recast classical mechanics as classical finite-state
dynamics, we can regard quantum mechanics as {\em generalizing}
classical mechanics---just as quantum computing generalizes classical
computing.  From this viewpoint, classical mechanical quantities are
simple special cases of quantum ones.

\vspace{3em}
\noindent
{\bf Acknowledgments} I thank Gerald Sussman, Jeffrey Yepez, Samuel
Braunstein, Lorenzo Maccone, Charles Bennett, Lev Levitin, Tom Toffoli,
Carlton Caves and William Unruh for their comments.

\setcounter{secnumdepth}{1}
\begin{appendix}

\section{Numerical Tests}\label{sec.tests}

A Mathematica notebook, containing code and results for numerical
experiments that confirm and extend the energy bound analysis above,
and generate the graphs in the figures above and below, is available
online \cite{code}.

The fundamental minimization problem outlined above requires
determination of non-negative coefficients $\av{a_n}^2$ that
minimize \eqn{dnu} while satisfying \eqn{orthogt} for a given set of
separations in time between distinct states, using the spectrum
\eqn{nun}.  Both the objective function \eqn{dnu} (raised to the
$M$ power) and the constraints \eqn{orthogt} are linear
combinations of the coefficients, so given a set of separations
between distinct states, we can find the global minimum to
arbitrary accuracy using linear optimization (linear programming).
We take separations to be integers, allowing us to deal with only a 
finite number of $\av{a_n}^2$ in our minimization: if both the
total period $T$ and the time $t$ are integers, then $T$ is the
number of distinct phases possible in the
constraints \eqn{orthogt}.  Using more than $T$ consecutive
$\av{a_n}^2$ with a given $\alpha$ would increase the frequency
moments \eqn{dnu} without allowing any new constraints.  Large
integer $T$ allows as much resolution in $t/T$ as desired.

\begin{figure}[H]{$\begin{array}{r}
 \includegraphics[width=7in]{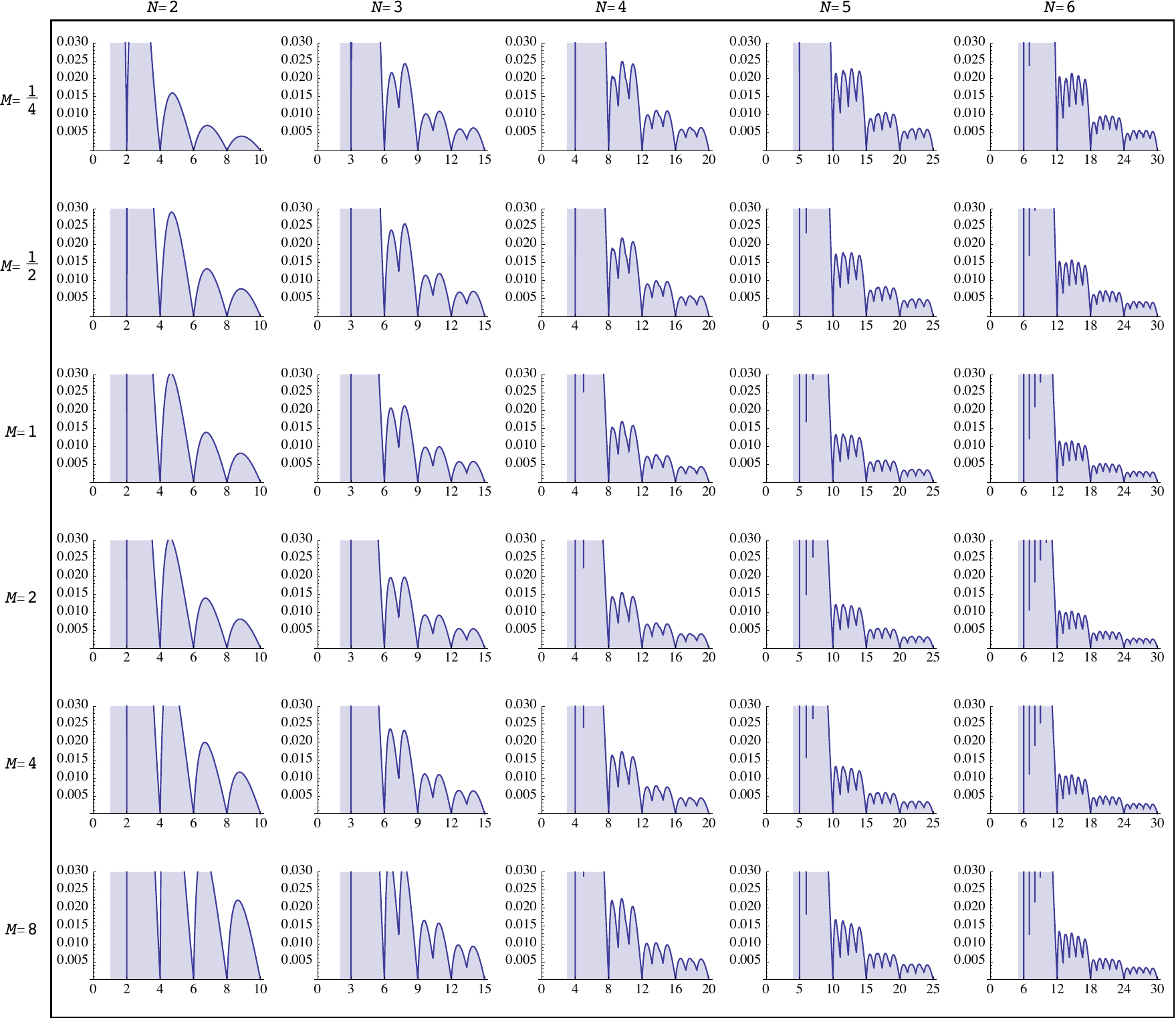}\\
 {\begin{minipage}[c]{7in}\centering\huge$\vdots$\end{minipage}}\\[.33in] \includegraphics[width=7.02in]{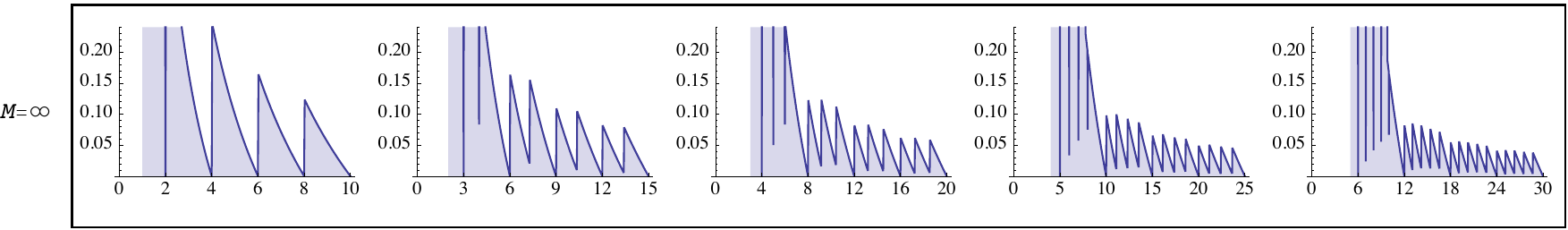}\end{array}$
 } \caption{{\bf Moments about a minimum frequency for a constant rate
 portion of evolution.}  As in Figure~\ref{fig.epart}, the graph at row
 $M$ and column $N$ shows the minimum value of
 $\tau\,\mom{\nu-\nu_0}{M}$ for each choice of maximum period $T$ for
 an evolution that includes $N$ distinct states separated by $N-1$
 equal intervals $\tauavg$, with the horizontal axes labeled with
 $T/\tauavg$.  For easier comparison, the $T=N\tauavg$ bound
 $f_{\nu_0}(M,N)$ is subtracted from each value plotted.}
\label{fig.epart-min-survey}
\end{figure}

\begin{figure}[H]{$\begin{array}{r}
  \includegraphics[width=7in]{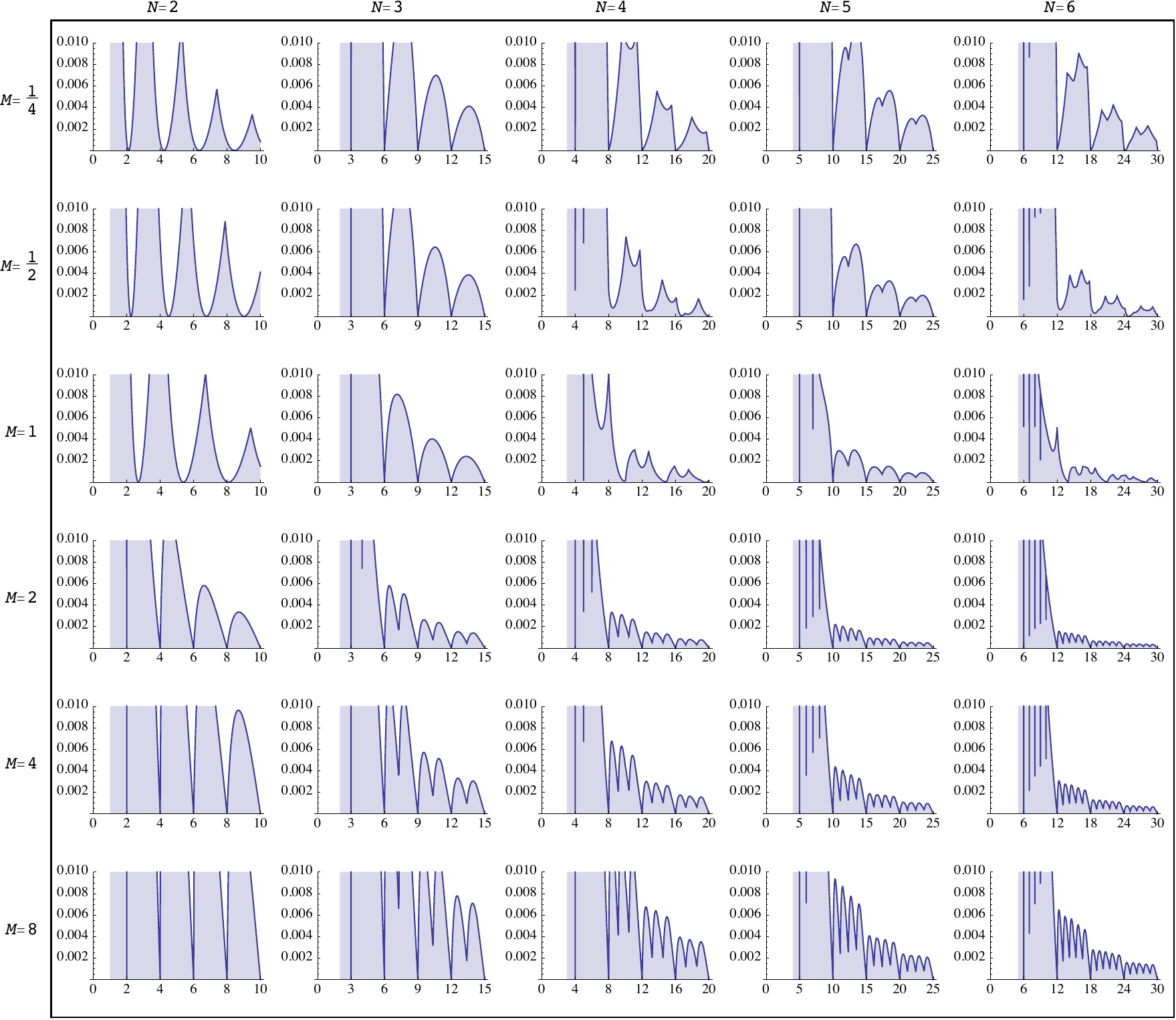}\\
  {\begin{minipage}[c]{7in}\centering\huge$\vdots$\end{minipage}}\\
  [.33in] \includegraphics[width=7.015in]{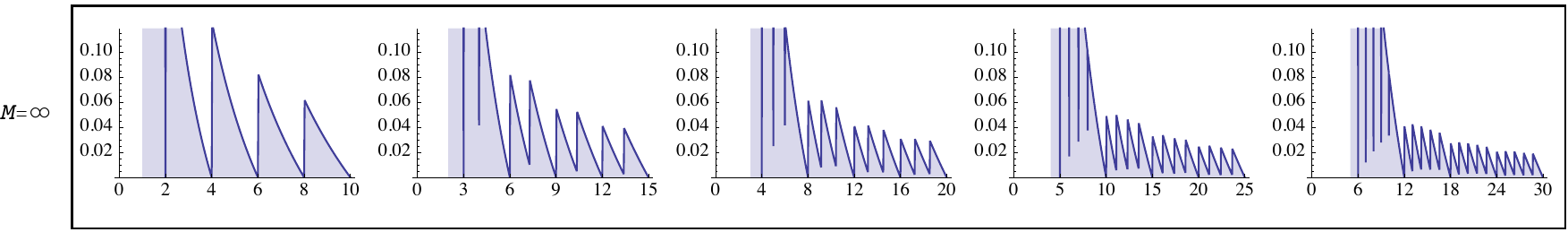} \end{array}$} \caption{{\bf
  Moments about the mean frequency for a constant rate portion of
  evolution.} The graph at row $M$ and column $N$ shows the minimum
  values of $\tau\,\mom{\nu-\bar\nu}{M}$ for each choice of maximum
  period $T$ for an evolution that includes $N$ distinct states
  separated by $N-1$ equal intervals $\tauavg$, with horizontal axes
  labeled with $T/\tauavg$.  In each graph the global minimum is
  subtracted, which is equal to the $T=N\tauavg$ bound
  $f_{\bar\nu}(M,N)$ except for some $M<2$ with $N$ even.  The case
  $M=1$ and $N=2$ is shown in detail in
  Figure~\ref{fig.mean-bound}.}
\label{fig.epart-mean-survey}
\end{figure}

\begin{figure}[H]{ 
\begin{minipage}[t]{\textwidth} \hspace{.05in}\includegraphics[width=2.3in]{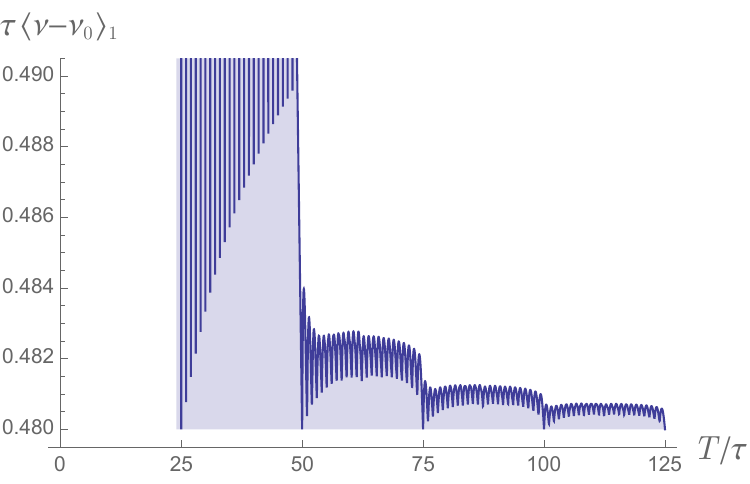}\hspace{.13in}\includegraphics[width=2.24in]{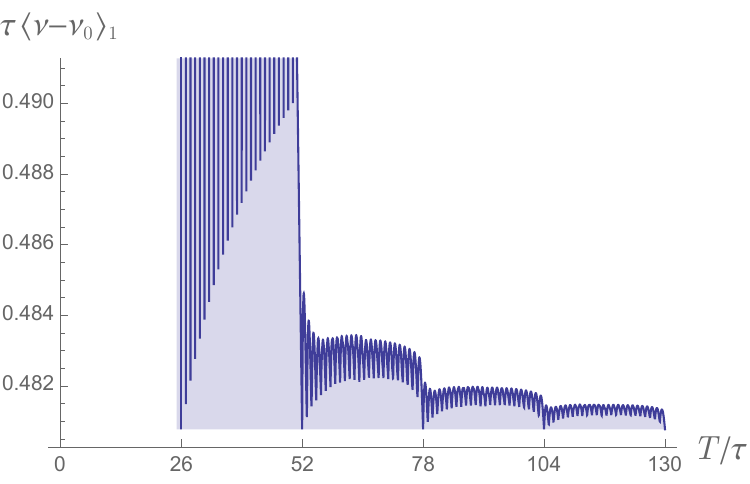}\hspace{.09in}\includegraphics[width=2.3in]{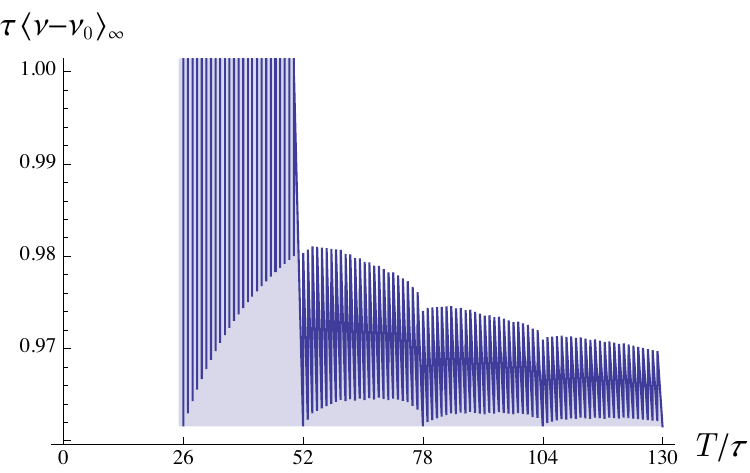}\end{minipage}\\[1em] 
\begin{minipage}[t]{\textwidth}   \includegraphics[width=2.355in]{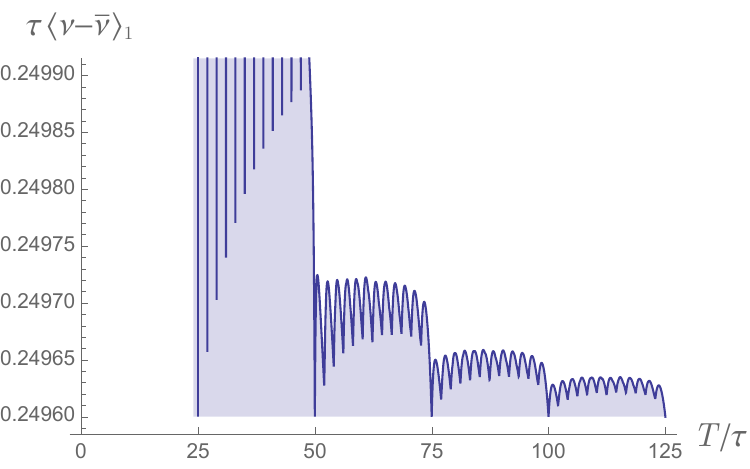}\hspace{.098in}\includegraphics[width=2.3in]{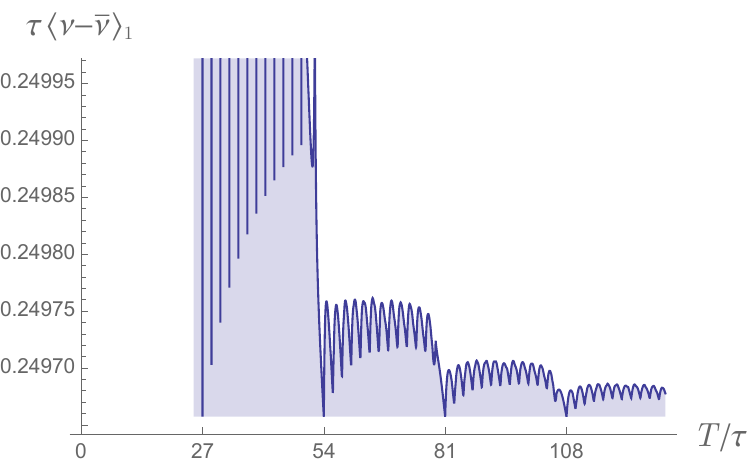}\hspace{.1in}\includegraphics[width=2.3in]{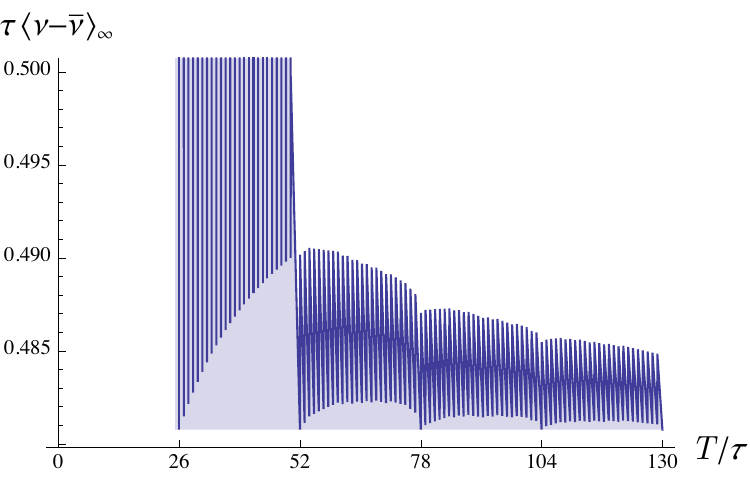}\end{minipage}}
  \caption{{\bf Minimizing moments for constant rate portion of
  evolution, using larger $\bm N$.}  Minimum values are computed
  numerically for $\tauavg\,\mom{\nu-\nu_0}{M}$ (first row) and
  $\tauavg\,\mom{\nu-\bar\nu}{M}$ (second row) as we vary the maximum
  period $T$, for an evolution that includes $N$ distinct states
  separated by $N-1$ equal intervals $\tauavg$.  The first column has
  $N=25$, the others $N=26$.  All global minima agree with the
  $T=N\tauavg$ bound $f_{\alpha}(M,N)$ except for the bottom middle
  case, as expected: here the smallest $\tauavg\,\mom{\nu-\bar\nu}{1}$
  agrees with $.249657= f_{\bar\nu}(1,27)$, rather than
  $.25=f_{\bar\nu}(1,26)$.}
\label{fig.epart-higher-n}  
\end{figure}

\begin{figure}[H]{ 
\begin{minipage}[t]{\textwidth} \includegraphics[height=2.175in]{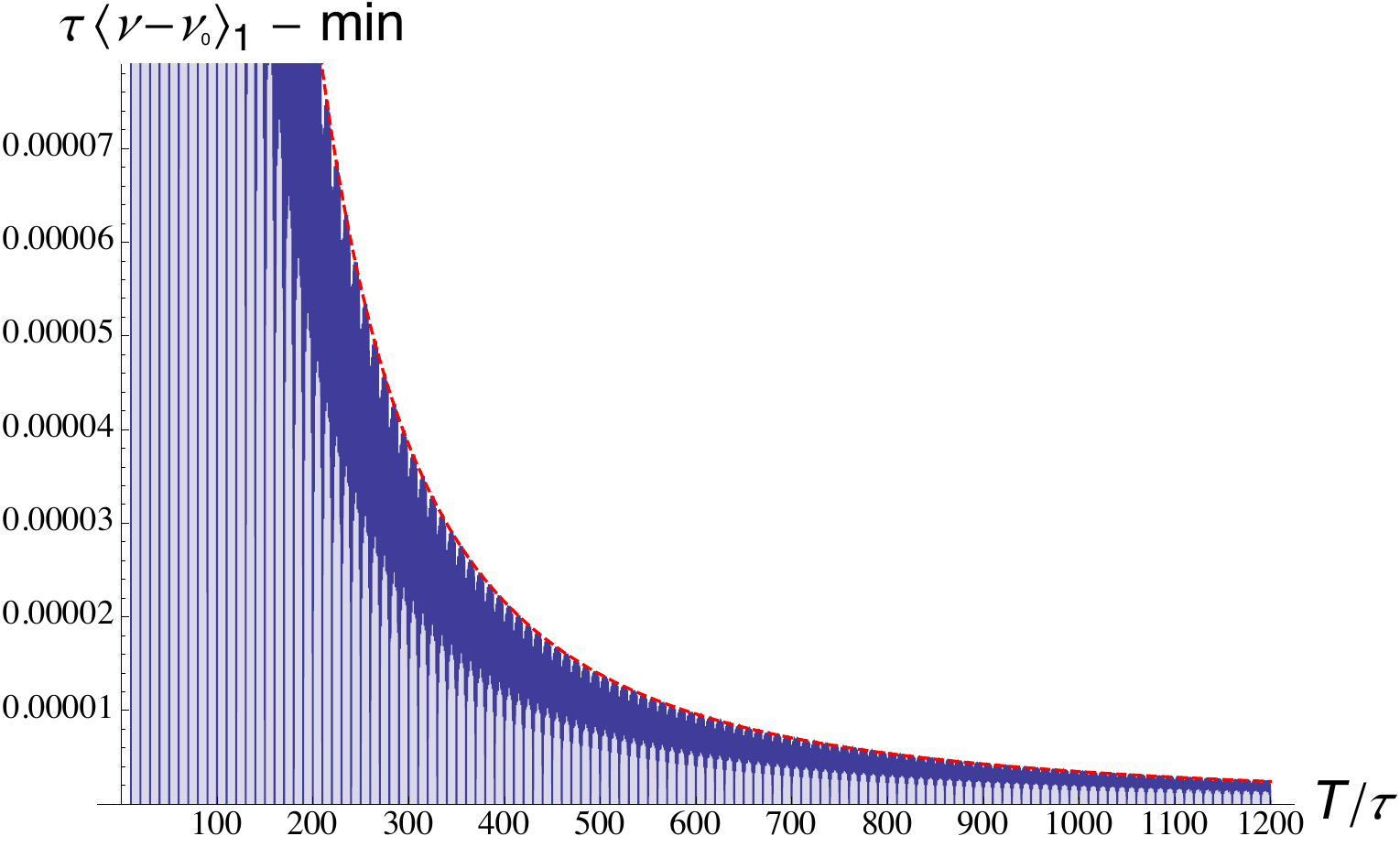}\includegraphics[height=2.175in]{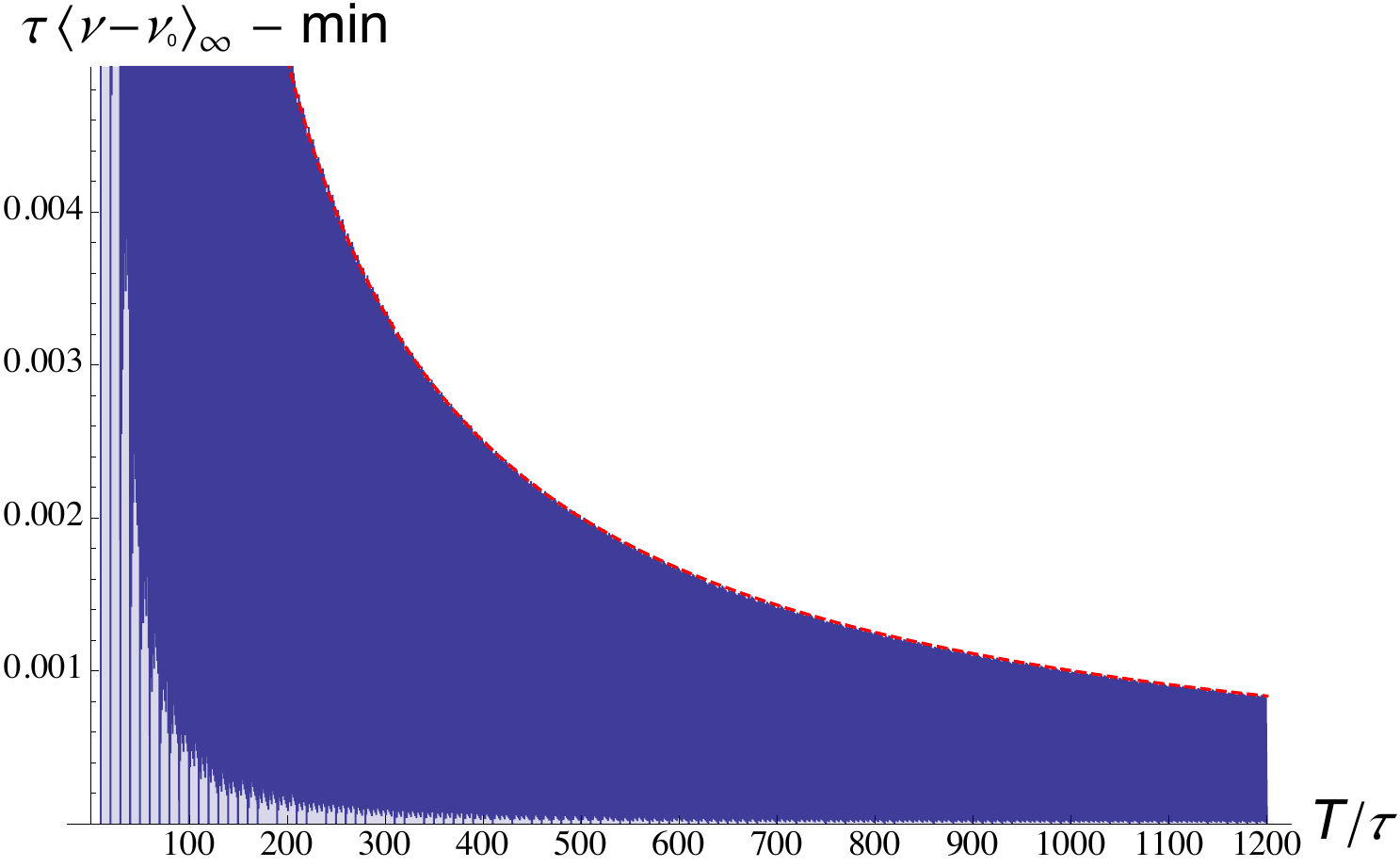} \end{minipage}}
  \caption{{\bf Asymptotic behavior of minima with constant rate
  portion of evolution.}  On the left we plot
  $\tauavg\,\mom{\nu-\nu_0}{1}$ minus its global minimum, for periodic
  evolutions of various lengths $T$ that include $N=10$ distinct states
  separated by $\tauavg$; similarly on the right for
  $\tauavg\,\mom{\nu-\nu_0}{\infty}$ with $N=10$. On the left, the red
  dashed boundary is the function $3.456 (T/\tauavg)^{-2}$. Other
  finite moments also fall asymptotically like $T^{-2}$.  On the right,
  the boundary is simply $(T/\tauavg)^{-1}$.  This is true of
  $\tauavg\,\mom{\nu-\nu_0}{\infty}$ for all $N$; the boundary for
  $\tauavg\,\mom{\nu-\bar\nu}{\infty}$ falls half as fast.}
\label{fig.epart-asymp}
\end{figure}

In surveying evolutions similar to Figure~\ref{fig.epart}, with a
portion constrained to go through $N$ distinct states with equal
separations $\tauavg$ (see Figures~\ref{fig.epart-min-survey}
through \ref{fig.epart-asymp}), the number of consecutive $\av{a_n}^2$
needed for large $T$ is only about $T/\tauavg$, rather than $T$.  This
is the asymptotically relevant bandwidth $1/\tauavg$ (discussed
earlier), divided by the spacing $1/T$ between allowed frequencies.
Neglecting the smallest possible values of $T$, which give minimum
moments too large to appear on our graphs, we find that in our tests,
enough $\av{a_n}^2$ for the largest $T$ is sufficient for all $T$.  Our
choice of $\tauavg$ sets the horizontal resolution of the
graphs---these examples use $\tauavg=43$.  For moments about a mean,
the position of the mean frequency relative to the other frequency
components makes a difference, so minimization for each choice of total
period $T$ involves searching a range of width $1/T$ for the $\alpha$
that minimizes $\mom{\nu-\alpha}{M}$.  For $M\ge 1$, the $\alpha$ found
is always the mean $\bar\nu$ of the minimizing state---except for $M=1$
with $T=N\tauavg$ and $N$ even, in which case all the $\alpha$ give the
same minimum.  For $0<M<1$ we must add a constraint to each
optimization problem, that the mean equals the $\alpha$ being tried.
Behavior similar to Figure~\ref{fig.epart} is seen for
$\tauavg\,\mom{\nu-\alpha}{M}$ for almost all $M$ (tested for $M$ up to
$1000$ and for $M=\infty$) and $N$ (tested up to $N=30$).  The only
exceptions are moments about $\bar\nu$ with $M< 2$ and $N$ even: in
some of these cases the intervals between the deepest local minima are
longer than $N\tauavg$, and in some cases the pattern of minima is less
regular.  Of course an estimate of the global minimum can always be
obtained by simply minimizing any case with large $T$.  In our tests
(see Figure~\ref{fig.epart-asymp}), the difference between local maxima
and the global minimum falls as $T^{-2}$ asymptotically for finite $M$,
and as $T^{-1}$ for $M=\infty$.  The latter result is implied by a
large-$T$ bandwidth bound of $1/\tauavg +1/T$: we need to round up the
asymptotically relevant bandwidth $1/\tauavg$ to an integer multiple of
$1/T$.  The $M=\infty$ graphs in all of these figures are obtained from
the band\-widths (or band\-widths above the mean) of states that
minimize $\tauavg \mom{\nu-\alpha}{M}$ for finite $M$.  For all data
shown, the minimizing band\-widths are independent of $M$ for $\nu_0$
and, for $M\ge 30$, for $\bar\nu$.

To verify that equal times between distinct states is optimal, we
performed experiments with unequal times.  For example, for
Figure~\ref{fig.unequal-times} we generated 12,000 sets of separations
stochastically; each set dividing a period $T$ into $N\le 12$
intervals; each set involving $N_\diff\le 4$ different interval lengths
separating adjacent distinct states.  Separations were integers between
1 and 100, except for five sets of separations near 1000.  For each set
of separations we used the total $T$ of the integer separations as the
number of consecutive $\av{a_n}^2$ to appear in the minimization.  We
did a fair sampling for each $N_\diff$, except that half of the choices
of number-of-repetitions of a length favored fewer lengths, and half of
the choices of a length favored the longer lengths. This helped fill
out the cases with lower minima using a short experiment---our original
experiment was completely unbiased and required a much larger number of
samples.  Similar experiments with other moments also verified equal
times as optimal.

\begin{figure}[H]{
  \includegraphics[width=2.2in]{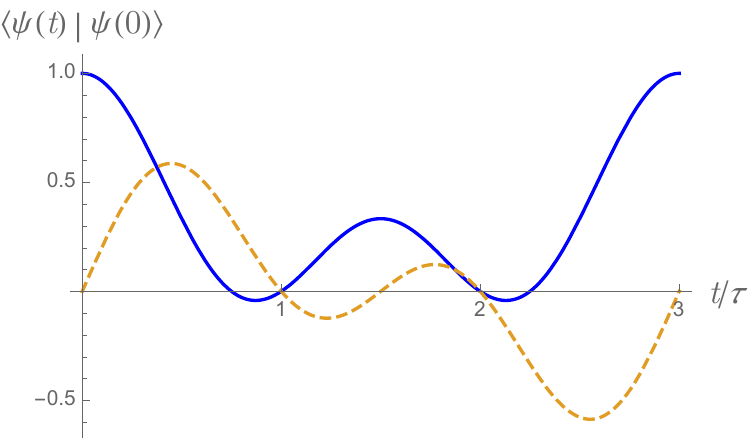} \hspace{.15in} \includegraphics[width=2.2in]{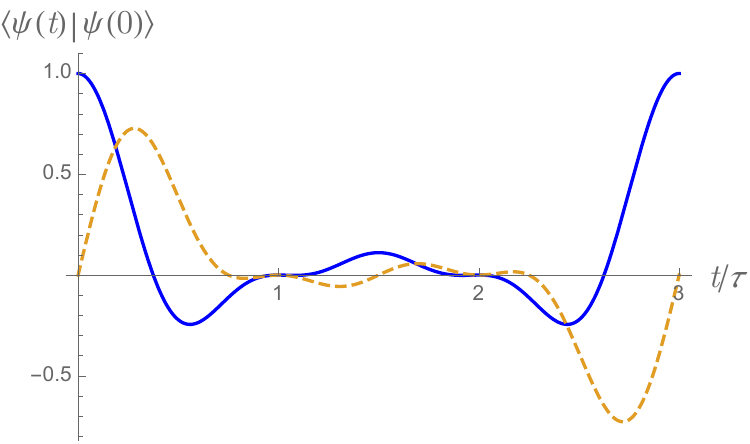}\hspace{.15in} \includegraphics[width=2.2in]{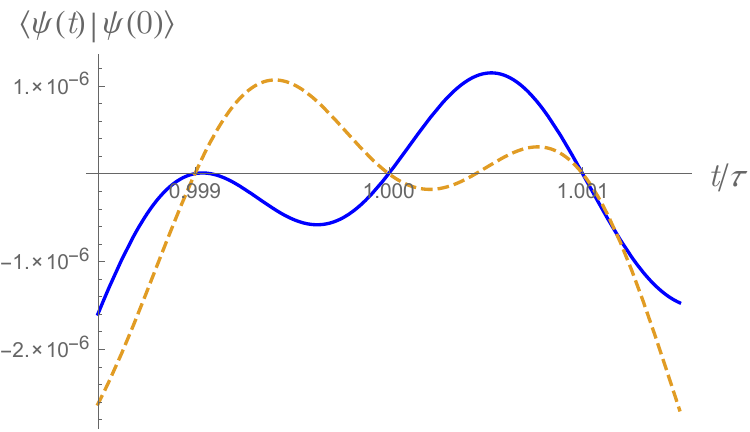}} \caption{{\bf
  Almost-equal separations.}  As in Figure~\ref{fig.overlap}, we show
  the real (solid) and imaginary (dashed) parts of the overlap function
  $\braket{\psi(t)}{\psi(0)}$ of Equation~(6), for $\ket{\psi(t)}$ that
  minimize $\mom{\nu-\nu_0}{1}$ for a periodic evolution of length $T$
  with $N$ distinct states. Left: All $N=3$ separations are of length
  $\tauavg=T/N$, and $\mom{\nu-\nu_0}{1}T=1$.  Middle: Separations
  differ by one part in $10^3$, and $\mom{\nu-\nu_0}{1}T\approx 2.001$.
  Right: Detail of flat region near $t/\tauavg =1$ from the middle
  graph.  If we make the separations more equal, the oscillation gets
  narrower and its amplitude smaller.  In the limit, only the extra
  constraint ``slope$\;=0$ at $t/\tauavg=1$ and 2'' keeps Middle
  distinct from Left, and $\mom{\nu-\nu_0}{1}T\to 2$.}
\label{fig.almost-equal-scaling}
\end{figure}
\begin{figure}[H]{ \hfill \mbox{
  \includegraphics[width=7in]{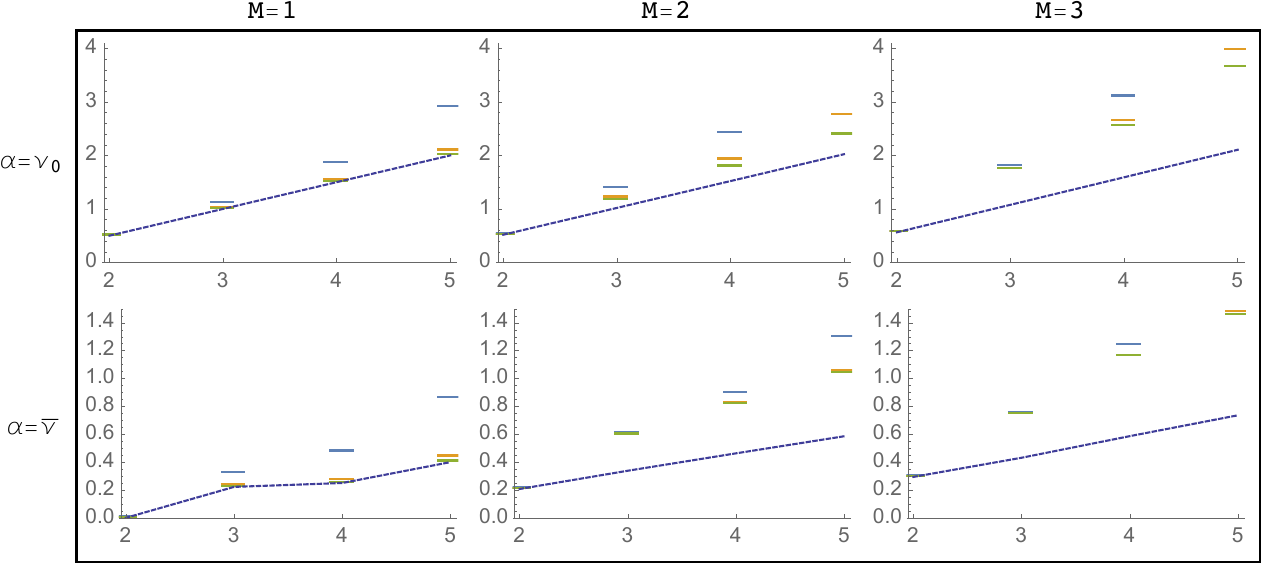}}\hfill} 
  \caption{{\bf
  Extra width required by almost-equal separations, above that for
  equal ones.}  All evolutions have period $T$.  Vertical axes show
  the extra minimum-width of $\mom{\nu-\alpha}{M}\,T$ for
  almost-equal separations, above the minimum needed for equal
  separations; horizontal axes show the number $N_\diff$ of
  different separations.  The dashed lines are theoretical curves
  that minimize the width assuming we impose just the usual
  equal-separation constraints, along with slope$\;=0$ constraints
  at the equal separations.  Triples of points correspond to
  separations that differ by one part in 10, $100$ or $1000$.  The
  theoretical bounds shown seem tight for $M=1$ or for
  $N_\diff=2$.}
\label{fig.almost-equal-survey}  
\end{figure}
\begin{figure}[H]{ \hfill \mbox{
  \includegraphics[width=2.8in]{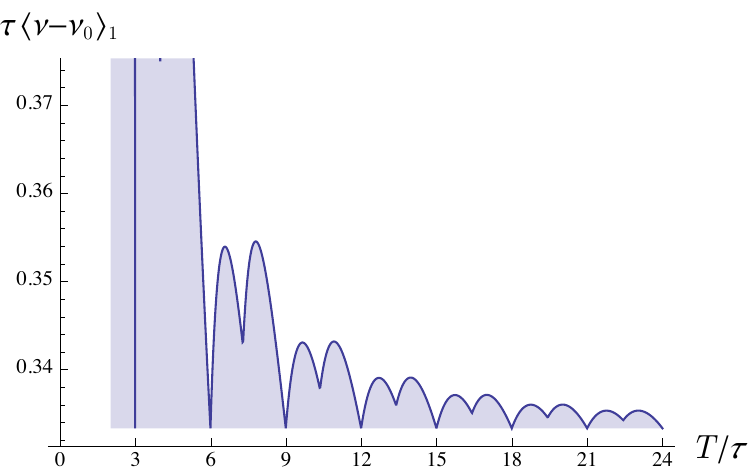}}\hspace{1in} \mbox{\includegraphics[width=2.8in]{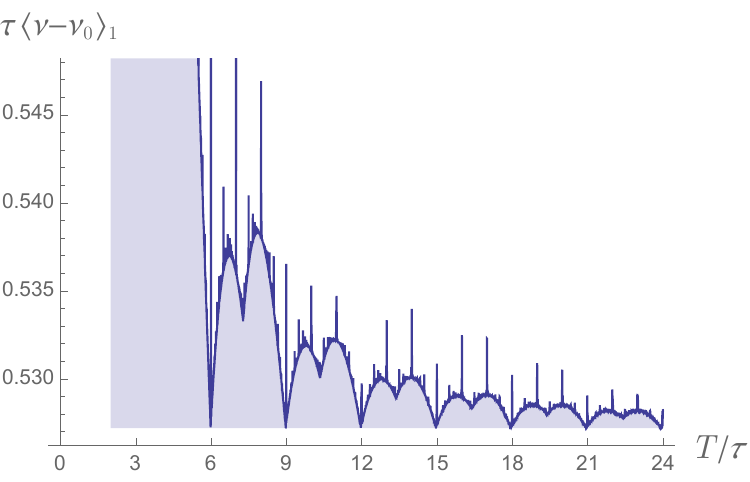}}\hfill} 
  \caption{{\bf
  Unequal separations in a portion of an evolution.} Left: Minimum of
  $\tauavg\,\mom{\nu-\nu_{0}}{1}$ for two equal separations between
  three distinct states.  Right: Two almost-equal separations require a
  larger $\tauavg\,\mom{\nu-\nu_0}{1}$.  The unequal separations used
  here differ by one part in $10^3$, and the spikes are not numerical
  artifacts.  With equal separations the minimum is 1/3; with the given
  unequal separations the minimum is about .527.  The range shown on
  the right is only half that on the left.}
\label{fig.almost-equal-comparison}
\end{figure}
\begin{figure}[H]{ \hfill \mbox{
    \includegraphics[width=2.1in]{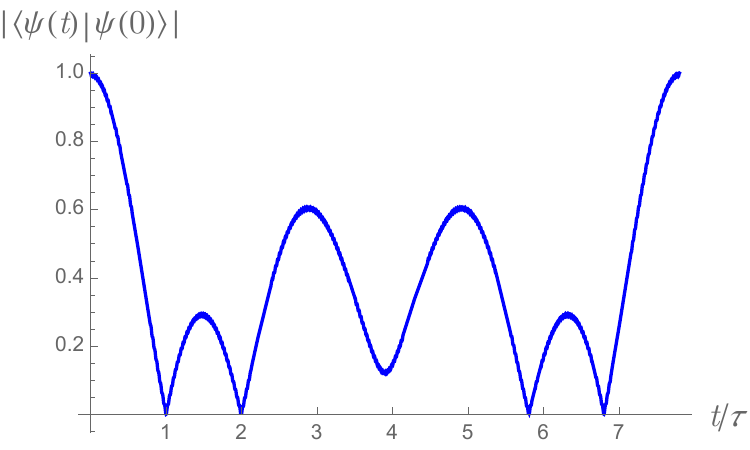}}\hspace{.3in} \mbox{\includegraphics[width=2.1in]{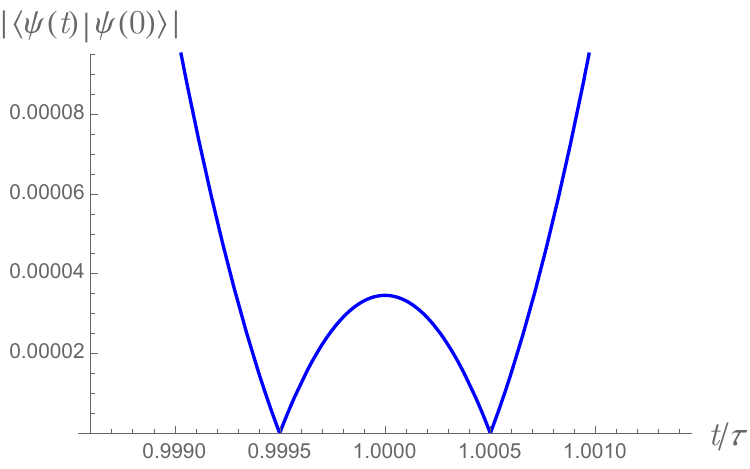}}\hspace{.3in} \mbox{\includegraphics[width=2.1in]{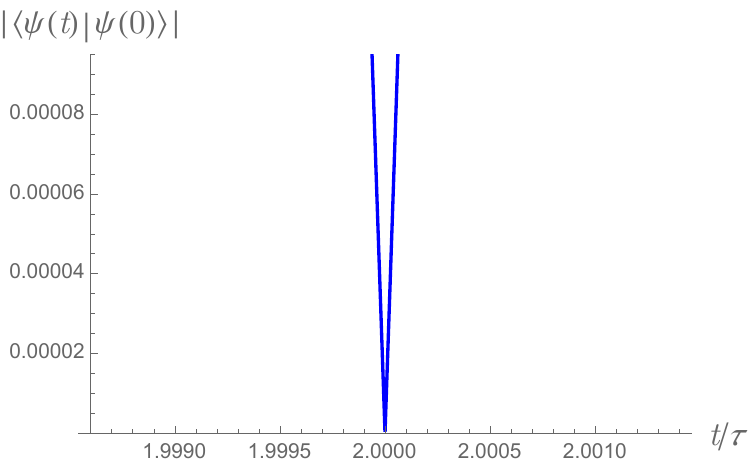}}\hfill}
  \caption{{\bf Satisfying almost-equal orthogonality constraints.}
  For 
  the computation shown in Figure~\ref{fig.almost-equal-comparison} (right), we look in
  detail at a particular value of the maximum period: for
  $T=7.8\,\tauavg$ we plot the magnitude of the overlap function
  $\braket{\psi(t)}{\psi(0)}$ of Equation~(6) using the
  coefficients $\av{a_n}^2$ that minimize $\mom{\nu-\nu_0}{1}$.
  Left: Full-scale behavior.  Middle: Detail near $t=\tauavg$.
  Right: Detail near $t=2\tauavg$.  The full scale graph depends
  strongly on our choice of $T$, but the detail graphs near
  $\tauavg$ and $2\tauavg$ don't.}
\label{fig.almost-equal-overlap}
\end{figure}

Since the dashed boundary in Figure~\ref{fig.unequal-times} is formed
by evolutions with almost-equal separations, we investigated those
cases extensively (see Figures~\ref{fig.almost-equal-scaling}
through \ref{fig.almost-equal-overlap}). As unequal separations
converge towards equal ones, the requirement that arbitrarily-close
points of the overlap must be zero contribute additional
slope-constraints on top of the equal-separation constraints, as shown
in Figure~\ref{fig.almost-equal-scaling}.  The theoretical curves
(dashed lines) in Figure~\ref{fig.unequal-times} and
Figure~\ref{fig.almost-equal-survey} were obtained by minimizing the
equal separations cases, with slope$\;=0$ constraints added at the
equal separations.  The triples of data points plotted for each $N$ in
Figure~\ref{fig.almost-equal-survey} used separations differing by one
part in 10, 100, and 1000.  For each minimization, we let the number of
frequency components equal the total integer period $T$.  The
minimizations about the mean for $N_\diff = 5$ used 600 decimal digits
of precision.  As is evident in the figure, the improvement in the
minimum from using smaller and smaller relative differences diminishes
rapidly.  There are similarly diminishing returns from using very large
numbers of $\av{a_n}^2$ with almost-equal separations.  A minimization
of $\mom{\nu-\nu_0}{1}\,T$ (not included in the figures) for $N=4$
different separations that differ from one another by only one part in
$10^6$, using 300,000 consecutive $\av{a_n}^2$, exceeded the
difference$\,\to\!0$ limiting value by only about two parts in $10^5$.
This was mostly accounted for by three very high frequency
components. The only other non-zero coefficients were $a_{14}$ and
below.  Figures~\ref{fig.almost-equal-comparison}
and \ref{fig.almost-equal-overlap} show an example of a portion of
evolution with separations that differ from each other by about one
part in $10^3$.
\vspace*{-.7em}

\onecolumngrid \newpage

\nolinenumbers

\setlength{\fboxsep}{.1pt} \setlength{\fboxrule}{.1pt}
\newcommand{\M}{{\scalebox{.65}{$M$}}}


\vspace{-.1em}

\newpage

\newpage

\newpage

\begin{figure}[H]{ \hfill \mbox{
    \includegraphics[width=2.1in]{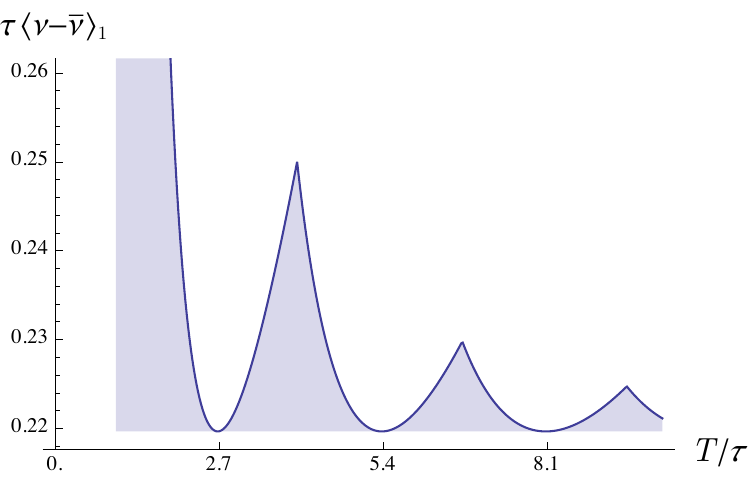}}\hspace{.3in} \mbox{\includegraphics[width=2.1in]{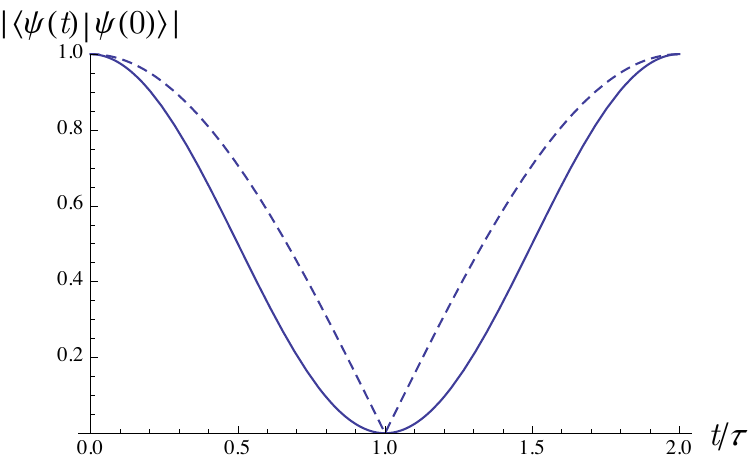}}\hspace{.3in} \mbox{\includegraphics[width=2.1in]{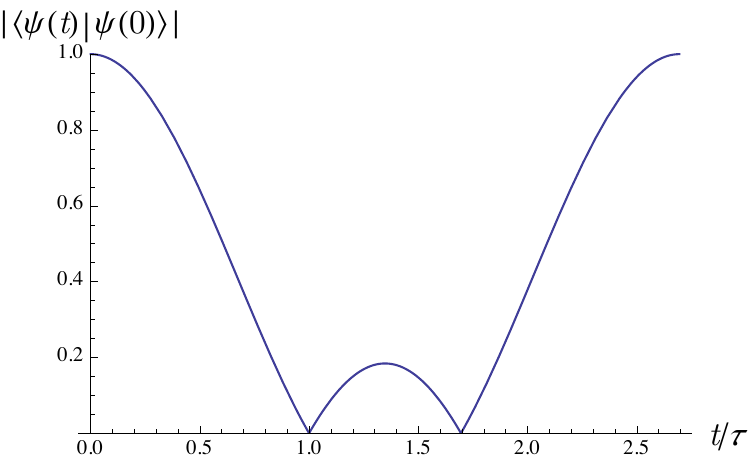}}\hfill}
  \caption{{\bf First absolute moment about the mean with two distinct states.}
  Left: If the two distinct states are separated by an interval
  $\tauavg$, the global minimum of $\tauavg\,\mom{\nu-\bar\nu}{1}$ is
  about $.22$, and repeats whenever the maximum period $T$ of the
  evolution is an integer multiple of approximately $2.7\tauavg$.
  Middle: At $T=2\tauavg$, a three-frequency-wide state (solid)
  achieves the same minimum $\tauavg\,\mom{\nu-\bar\nu}{1}$ as a
  minimum-width two-frequency state (dashed).  Right: The state that
  achieves the first global minimum at $T\approx 2.7\tauavg$ uses three
  frequencies.  Knowing this, we can determine the minimum analytically
  by solving a transcendental equation: $f_{\bar\nu}(1,2)\to u/2\pi$
  where $u\approx 1.3801$ is a root of $u\sin(u+\sqrt{u^2 -1})=1$.}
\label{fig.mean-bound}
\end{figure}  

\newpage\twocolumngrid

\section{Random hamiltonians}\label{sec.rand-h}

\begin{figure}[t]{ \hfill \mbox{
    \includegraphics[width=2.8in]{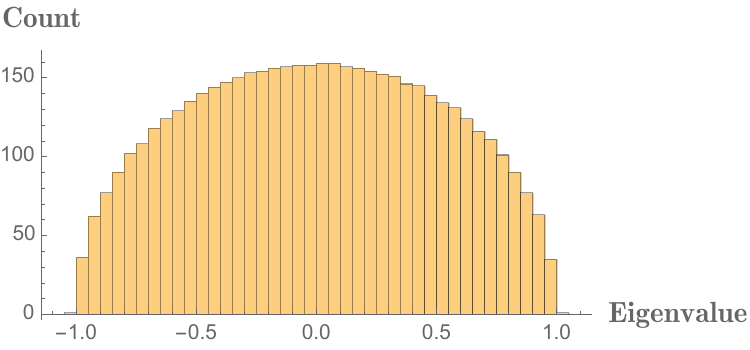}}\hfill} \caption{{\it
  Distribution of eigenvalues for a $5000\times 5000$ hermitian matrix
  with random entries.}  The matrix has been normalized to give a fixed
  range of eigenvalues.  Here the range is divided evenly into bins,
  and the number of eigenvalues in each bin is plotted. The shape is
  semi\-circular for almost any kind of randomness, if the width is
  plotted as twice the height.}
\label{fig.semi-circle}
\end{figure}

Wigner \cite{wigner55} observed that one can predict aspects of the
behavior of complicated dynamics by studying the distribution of
eigenvalues of hamiltonian matrices with random entries---a novel kind
of statistical mechanics \cite{guhr}.  Here we review how eigenvalue
statistics lead to a universal law governing how soon random
hamiltonian evolution reaches a state orthogonal to a generic initial
state \cite{rmt}.

The property of random hamiltonians that we use is illustrated in
Figure~\ref{fig.semi-circle}.  The histogram shows the number of
eigenvalues that fall in equal-sized ranges for a $5000\times 5000$
random hermitian matrix.  The semi\-circular shape is universal,
independent of the details of the randomness \cite{erdos}.  In the case
shown, the hermitian matrix was constructed by adding a random complex
matrix to its conjugate transpose, and then normalized by dividing by
half-the-width of its eigenvalue range.  The entries in the complex
matrix had real and imaginary parts chosen uniformly between $-1/2$ and
$+1/2$.

For an initial state $\psi(0)=\sum a_n \ket{E_n}$, the overlap with
$\psi(t)$ depends only on the squared magnitudes $\av{a_n}^2$ and the
energy eigenvalues $E_n$. If the initial state is randomly chosen, the
eigenvalues in each range appear in the overlap about as frequently as
they do in the semi\-circular distribution, so we can exactly compute
the overlap in the infinite dimensional limit.  For eigenvalues ranging
from $-E$ to $+E$, letting $h=2$, this has the universal form
\begin{align}\label{eq.bessel}
  \braket{\psi(t)} {\psi(0)} &=\int_{-1}^{1} e^{ i \pi E u
 t}\,\frac{2}{\pi}\sqrt{1-u^2} \,du \nonumber \\[.5em] &= {{2J_1(\pi E
 t)}\over \pi E t}\;,
\end{align}
where $J_1$ is a Bessel function.  The factor $\sqrt{1-u^2}$ is the
height of a radius-one semicircle at position $u$. Multiplying this by
$2/\pi$ gives the probability density for a semi\-circular
distribution.  If we take the lowest eigenvalue to be zero, $E$ is the
average energy.

\section{Finite-state classical mechanics}\label{sec.cm}

Classical lattice gases, such as the Ising Model, have long played an
important role in statistical mechanics \cite{ruelle}.  Lattice gases
have also been used to model dynamics.  The simplest way to {\em
exactly} map continuous quantities in classical mechanics onto a
finite-state lattice is to first construct a continuous evolution that
has discrete properties at integer times---and then just model the
integer time
behavior \cite{bbm,fhp,super,salem,rothman,ssm,crystalline,ideal-energy}
({\em cf.} \cite{marsden,bahr}).  For example, model a gas of idealized
particles that, if started exactly in one of a finite number of
configurations of positions and momenta, is always found in one of
these configurations at integer times.  As long as we only constrain
the initial state to be discrete, and not the dynamical law itself,
conservations still follow from continuous symmetries of the
lagrangian.

\begin{figure}[t]{ \hfill \mbox{
    \includegraphics[width=2.6in]{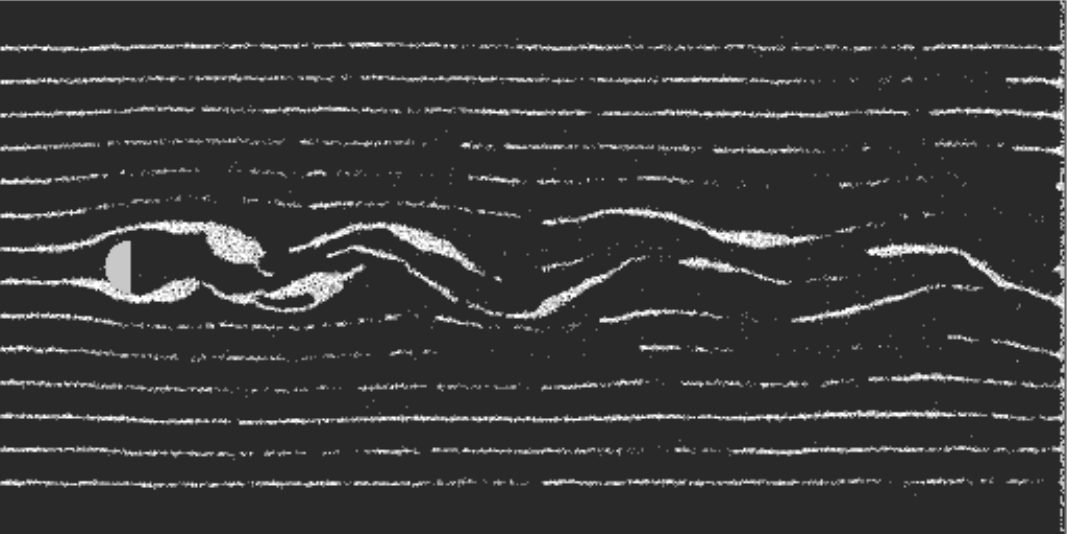}}\hfill} \caption{{\it Realistic
  hydrodynamic flow in a finite-state lattice gas.}  We first construct
  an idealized continuous dynamics where, if particles start at lattice
  locations with a discrete set of velocities, they are always found at
  lattice locations at integer times. We then simulate just the integer
  time behavior.}
\label{fig.fhp}
\end{figure}

\begin{figure}{ \hfill \mbox{
    \includegraphics[width=2.3in]{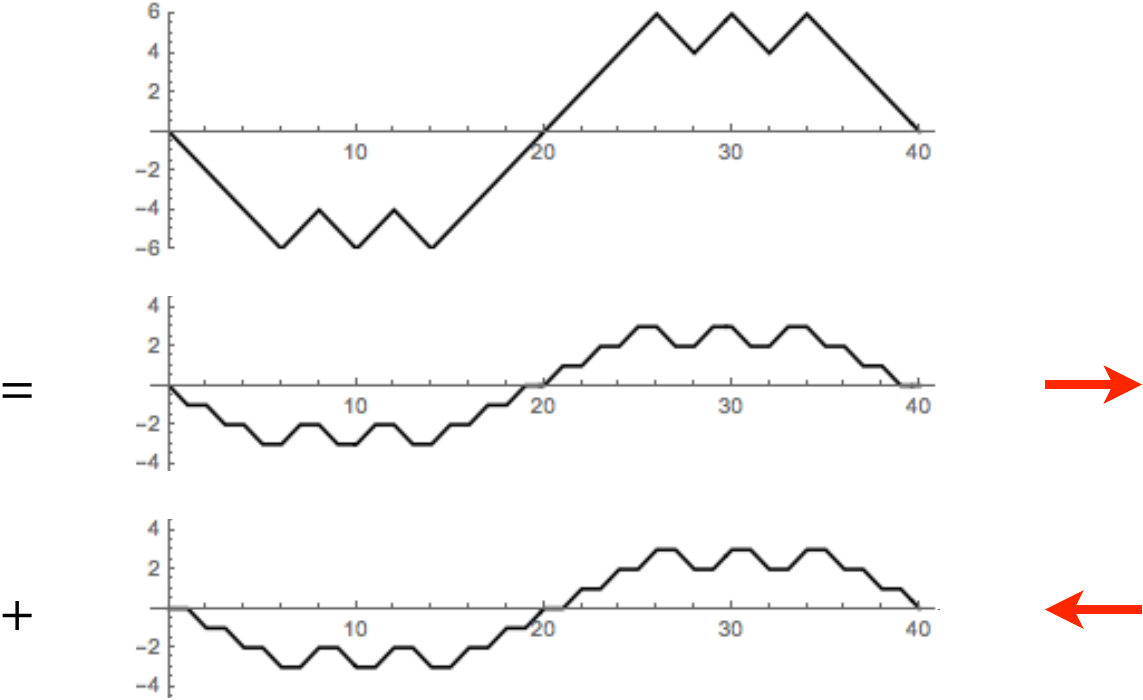}}\hfill} \caption{{\it
    Continuous wave-equation in a finite-state lattice dynamics.}  One
    dimensional wave behavior is a superposition of a right-going wave
    and a left-going wave.  If the moving waves have discretely
    constrained shapes, then at integer times they exactly match the
    behavior of a simple local lattice dynamics.}
\label{fig.waves}
\end{figure}

Figure~\ref{fig.fhp} illustrates a simple discrete-velocity lattice
gas, constructed by restricting the initial positions and momenta of an
idealized classical mechanical gas so that particles are always found
on a triangular lattice at integer times.  At large scales, and with
relatively slow flow rates, the lattice-scale constraints on the
initial state become invisible and the flow is hydrodynamic, with full
rotational symmetry \cite{fhp}.  For visualization, a second lattice
gas that follows the flow was added (shown in white).  Similar models
in three dimensions enable realistic hydrodynamic simulations of
complex fluids \cite{rothman}.

Figure~\ref{fig.waves} provides a second example, illustrating how the
continuous wave-equation can be simulated by simple finite-state
lattice dynamics.  Again we start with continuous motion---right and
left going waves---and impose constraints on the initial state so that,
at integer times, only configurations from a finite set are possible.
The integer time behavior thus samples the continuous wave equation,
albeit microscopically there are constraints on the detailed initial
shape of the wave.  This mechanism was actually discovered
experimentally, in simple reversible cellular automata models of
physics such as energy conserving Ising dynamics \cite{crystalline}.
Exactly invertible lattice simulations of the wave equation can be
implemented in any number of dimensions \cite{joe}.  Similar
examples \cite{ideal-energy} sample continuous relativistic dynamics.

\section{Interpolated evolution}\label{sec.interpol}

\begin{figure}[t]{ \hfill \mbox{
    \includegraphics[width=2.5in]{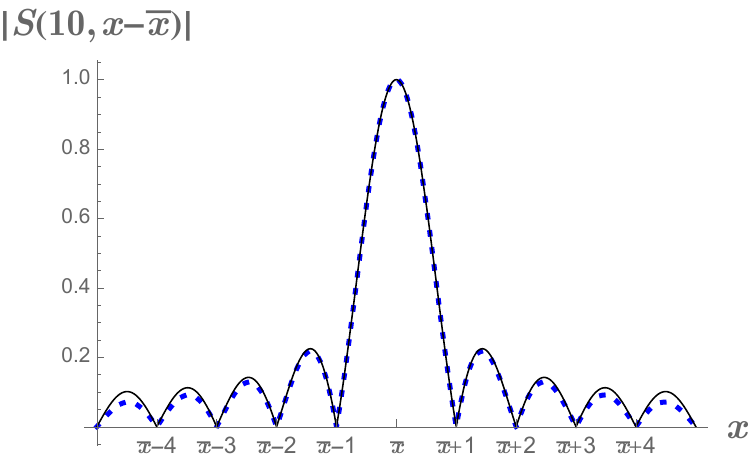}}\hfill\hfill} \caption{{\em
  Magnitude of periodic sampling function.}  One period of
  $\left\lvert{S(10,x-\bar{x})}\right\rvert$ is shown (solid), centered
  at position $\bar{x}$.  Like the magnitude of the usual sinc sampling
  function (shown dotted), it is zero at integer separations from
  $\bar{x}$ (but is periodic).}
\label{fig.s20}
\end{figure}

The quantum formalism provides a natural mechanism for turning a
discrete evolution into a continuous one \cite{emulation}.  If, for
example, we define a unitary $\op{U}_\tau$ that performs a discrete
logic operation on qubits in a fixed time $\tau$, we can find a
hamiltonian $\op{H}$ that generates $\op{U}_\tau$ in time $\tau$:
\begin{equation}\label{eq.utau}
\op{U}_\tau=e^{-2 \pi i \op{H} \tau / h}\;.
\end{equation}
Any such $\op{H}$ defines an evolution not only at intervals $\tau$,
but also for any other time interval $t$, with $\op{U}_t = e^{-2 \pi
i \op{H} t / h}$.

Constructing an $\op{H}$ from a discrete evolution is a kind of
interpolation \cite{interpolation} that turns a set of samples at
discrete times into a continuous evolution.  To illustrate the
connection to classical interpolation, consider a simple classical
finite-state evolution: at integer multiples of time $\tau$, a single 1
appears at consecutive integer positions of a 1D periodic space of
width $N$.  All other integer positions contain 0's at these times.  We
construct an $\op{H}$ that implements this discrete logical shift as
part of a continuous-time quantum evolution that achieves the average
energy bound \eqn{et} on distinct state change.

We take the $N$ distinct logical configurations as basis states.  Let
$\ket{n}$ be the state where the 1 is at position $n$, and $\ket{n+1} =
\op{U}_\tau \ket{n}$. Define another set of basis vectors $\{\ket{k}\}$
as the fourier transform of the position basis $\{\ket{n}\}$:
\begin{equation}\label{eq.Em}
  \ket{k} = {1\over \sqrt{N}}\sum_{n=0}^{N -1} e^{+2\pi i n
    k/N}\,\ket{n}\;\;
\end{equation}
for integers $k\in\left[ 0,N-1 \right]$.  The inverse transform is then
\begin{equation}\label{eq.inv-trans}
  \ket{n} = {1\over \sqrt{N}}\sum_{k=0}^{N -1} e^{-2\pi i n
    k/N}\,\ket{k}\;.
\end{equation}
Define $\op{H}$ by letting each $\ket{k}$ be an energy eigenstate, with
energy eigenvalue $E_k = k h/ N\tau$.  Then if $\op{U}_t=e^{-2\pi i \op{H} t/h}$,
\begin{align}\label{eq.step}
  \op{U}_\tau\,\ket{n} &= {1\over \sqrt{N}}\sum_{k=0}^{N -1} e^{-2\pi i
    (n+1) k/N}\,\ket{k} \nonumber \\[.3em] &= \ket{n+1}\;,
\end{align}
which is the desired discrete evolution.  We see from \eqn{inv-trans}
that $\ket{n}$ is an equally weighted superposition of equally
separated energy eigenstates, so this evolution achieves the average
energy bound on distinct state change.

Given $\op{H}$, the state at any continuous time $t$ can be expressed
in terms of the integer-position basis states.  Starting with the
particle in a basis state at position 0,
\begin{align}\label{eq.t-from-n}
  \op{U}_t\,\ket{0} &= {1\over \sqrt{N}}\sum_{k=0}^{N -1} e^{-2\pi i k
    t/N\tau }\,\ket{k} \nonumber \\[.3em] &={1\over N} \sum_{k=0}^{N
    -1} e^{-2\pi i k t/N\tau }\,\sum_{n=0}^{N -1} e^{+2\pi i n
    k/N}\,\ket{n} \nonumber \\[.3em] &=\sum_{n=0}^{N - 1}
    S(N,n-t/\tau)\:\ket{n}\;,
\end{align}
where
\begin{equation}\label{eq.S}
  S(N,u) = {1\over N} \sum_{k=0}^{N -1} e^{2\pi i k u/ N}\;
\end{equation}
is a periodic version of the sampling function $\sinc \pi u = \sin \pi
u\,/ \,\pi u$, which is the foundation of bandlimited interpolation
theory.  This is illustrated in Figure~\ref{fig.s20}.  In fact,
$S(N,u)\to e^{i\pi u}\,\sinc \pi u$ for $N\to\infty$.  For $t/\tau$ an
integer, only one configuration $\ket{n}$ in \eqn{t-from-n} has a
non-zero coefficient.  At all other times, $S(N,n-t/\tau)$ is not
centered at an integer, and all $\ket{n}$ have non-zero coefficients.

\section{Sampled evolution}\label{sec.sampled}

A continuous evolution is equivalent to a discrete one if the
continuous is just an interpolation of the discrete: then there is no
more information in the continuous than in the
discrete \cite{interpolation,emulation,kempf1,kempf2,succi,birula,meyer,yepez-path,yepez-gr,yepez,qsim,min-len,mauro,tsang,ideal-energy}.
This equivalence is the basis of sampling theory: a finite-bandwidth
periodic signal has only a finite number of terms in its Fourier
series, so once enough discrete samples of the signal have been taken
to determine the coefficients of all terms, there is no more
information in the signal.  The signal is effectively discrete.

The same logic applies to quantum systems: if the evolution is periodic
and uses only a finite range of energy frequencies, then knowing a
finite number of samples $\ket{\psi(t_k)}$ of the continuous (vector)
signal $\ket{\psi(t)}$ determines the rest.  This implies that
finite-sized quantum systems with finite energy are effectively
discrete, since their evolution is arbitrarily close to one cycle of an
exactly recurrent evolution with finite
bandwidth \cite{qm-recurrence}.

As an illustration of sampling a quantum evolution \cite{emulation},
consider a scalar wavefunction $\psi(x,t)$ that has a periodic
evolution with period $T$, and a range of energy eigenfrequencies of
width $(K-1)/T$, with $E_0=0$.  Using units of time where $T=K$, we see
that the state at all times is just an interpolation of the state at
integer times:
\begin{equation}\label{eq.time-samples}
\textstyle
\psi(x,t) = \sum_{k=0}^{K-1} \psi(x,k)\: S(K,k-t)\;,
\end{equation}
where $S$ is the sampling function \eqn{S}.  This identity is obvious
if $t$ is an integer between $0$ and $K-1$, since then
$S(K,k-t)=\delta_{k,t}$.  Hence \eqn{time-samples} holds for all $t$,
since both sides add together the same $K$ Fourier components, and the
$K$ integer-$t$ cases determine all coefficients in the sum.

This identity applies to the discrete shift example of
Appendix~\ref{sec.interpol}, recast as a continuous shift of a
finite-bandwidth state: $\op{H}= v\op{p}_x$, with $v=1$, and using only
a finite range of momenta in a periodic space.  To have $N$ distinct
states requires $K\ge N$, and to achieve the minimum bandwidth we must
use an equal superposition of equally separated eigenstates, which is
just $\psi(x,t)=S(N,x-t)$.  This evolution is a continuous
interpolation of a discrete (integer $x$ and $t$) binary shift.


\end{appendix}

\end{document}